\documentclass[fleqn]{article}
\usepackage{amsmath,amsfonts,amssymb,latexsym,cite}
\usepackage{graphicx}

\def\noi{\noindent}
\newcommand{\Title}[1]{\noi {{\Large\bf #1}}\\[1ex]}

\def\Aunames#1{\noi{\bf #1}}

\def\Addresses#1{\medskip\noi \protect
	\begin{description}\itemsep -3pt {\it #1} \end{description}}
\def\addr#1#2{\item[${}^{#1}$]{\it #2}}

\newcommand{\Abstract}[1]{\vskip 2mm \begin{center}
        \parbox{16.4cm}{\small\noi #1} \end{center}\medskip}

\def\email#1#2{\footnotetext[#1]{e-mail: #2}\addtocounter{footnote}{1}}

\def\nqq{\hspace*{-2em}}
\def\nhq{\hspace*{-0.5em}}

\def\cm{\hspace*{1cm}}

\def\ten#1{\mbox{$\cdot 10^{#1}$}}

\def\Jl#1#2{#1 {\bf #2},\ }
\def\ApJ#1 {\Jl{Astroph. J.}{#1}}
\def\CQG#1 {\Jl{Class. Quantum Grav.}{#1}}
\def\DAN#1 {\Jl{Dokl. AN SSSR}{#1}}
\def\GC#1 {\Jl{Grav. Cosmol.}{#1}}
\def\GRG#1 {\Jl{Gen. Rel. Grav.}{#1}}
\def\JETF#1 {\Jl{Zh. Eksp. Teor. Fiz.}{#1}}
\def\JETP#1 {\Jl{Sov. Phys. JETP}{#1}}
\def\JHEP#1 {\Jl{JHEP}{#1}}
\def\JMP#1 {\Jl{J. Math. Phys.}{#1}}
\def\NPB#1 {\Jl{Nucl. Phys. B}{#1}}
\def\NP#1 {\Jl{Nucl. Phys.}{#1}}
\def\PLA#1 {\Jl{Phys. Lett. A}{#1}}
\def\PLB#1 {\Jl{Phys. Lett. B}{#1}}
\def\PRD#1 {\Jl{Phys. Rev. D}{#1}}
\def\PRL#1 {\Jl{Phys. Rev. Lett.}{#1}}

\def\lal{&&\nqq {}}
\def\eq{Eq.\,}
\def\eqs{Eqs.\,}
\def\beq{\begin{equation}}
\def\eeq{\end{equation}}
\def\bear{\begin{eqnarray}}
\def\bearr{\begin{eqnarray} \lal}
\def\ear{\end{eqnarray}}
\def\earn{\nonumber \end{eqnarray}}

\def\nnn{\nonumber\\ \lal }

\def\yyy{\\[5pt] \lal }

\def\sequ#1{\setcounter{equation}{#1}}

\def\dst{\displaystyle}
\def\tst{\textstyle}
\def\fracd#1#2{{\dst\frac{#1}{#2}}}
\def\fract#1#2{{\tst\frac{#1}{#2}}}
\def\Half{{\fracd{1}{2}}}
\def\half{{\fract{1}{2}}}

\def\e{{\,\rm e}}
\def\d{\partial}

\def\diag{\mathop{\rm diag}\nolimits}

\def\const{{\rm const}}

\def\then{\ \Rightarrow\ }

\usepackage{color}

\def\rf{\eqref}
\def\eqn#1{Eq.\,\rf{#1}}

\def\mn{_{\mu\nu}}

\def\mN{_\mu^\nu}

\def\Z{{\mathbb Z}}
\def\cR{{\cal R}}

\def\wh{wormhole}
\def\whs{wormholes}
\def\bh{black hole}
\def\bhs{black holes}

\def\sph{spherically symmetric}
\def\ssph{static, spherically symmetric}
\def\asflat{asymptotically flat}
\def\asdS{asymptotically de Sitter}
\def\asAdS{asymptotically anti-de Sitter}
 
\def\KS{Kan\-tow\-ski-Sachs}

\def\Scz{Schwarz\-schild}

\begin{document}
\twocolumn[
\thispagestyle{empty}

\vspace{-5mm}

\Title{Wormholes with fluid sources: A no-go theorem and new examples}

\Aunames{K. A. Bronnikov$^{a,b,c,1}$,  K.A. Baleevskikh$^{b,2}$, and M.V. Skvortsova$^{d,3}$}

\Addresses{
\addr a  {Center of Gravitation and Fundamental Metrology,
       VNIIMS, Ozyornaya St. 46, Moscow 119361, Russia}
\addr b {Institute of Gravitation and Cosmology,  Peoples' Friendship University of Russia 
        (RUDN University), Miklukho-Maklaya St. 6, Moscow 117198, Russia}
\addr c {National Research Nuclear University MEPhI (Moscow Engineering Physics Institute),
       Kashirskoe highway 31, Moscow, 115409, Russia }
\addr d 
   {Peoples' Friendship University of Russia (RUDN University), Miklukho-Maklaya St. 6, Moscow 117198, Russia}
	}

\Abstract
   {For static, spherically symmetric space-times in general relativity (GR), a no-go theorem is proved: 
    it excludes the existence of wormholes with flat and/or AdS asymptotic regions on both sides of 
    the throat if the source matter is isotropic, i.e., the radial and tangential pressures coincide. 
    It explains why in all previous attempts to build such solutions it was necessary to introduce 
    boundaries with thin shells that manifestly violate the isotropy of matter. 
    Under a simple assumption on the behavior of the spherical radius $r(x)$, we obtain a number
    of  examples of wormholes with isotropic matter and one or both de Sitter asymptotic regions, 
    allowed by the no-go theorem. We also obtain twice asymptotically flat wormholes with 
    anisotropic matter, both symmetric and asymmetric with respect to the throat, under the assumption
    that the scalar curvature is zero. These solutions may be on equal grounds interpreted as those of GR 
    with a traceless stress-energy tensor and as vacuum solutions in a brane world. For such \whs, the
    traversability conditions and gravitational lensing properties are briefly discussed. 
    As a by-product, we obtain twice asymptotically flat regular black hole solutions with up to four 
    Killing horizons. As another by-product, we point out intersection points in families of integral curves 
    for the function $A(x) = g_{tt}$, parametrized by its values on the throat.  
   }

] %%%%%%%%%%%%%%%%%%%%%%%%%%%%%%
\email 1 {kb20@yandex.ru}
\email 2 {baleevskikh.k@gmail.com}
\email 3 {milenas577@mail.ru}

%=========================
\section{Introduction}

  Wormholes as two-way tunnels or shortcuts between different universes or different, otherwise
  distant regions of the same universe are at present a well-known and widely discussed subject,
  see, e.g., \cite{flamm, e-rosen, wheeler, viss-book, thorne, BR-book, lobo-rev, BS-extra}. 
  Wormholes are of interest not only as a perspective ``means of transportation'' but also as possible 
  time machines or accelerators \cite{thorne, BR-book}. The simplest \wh\ geometry is \ssph,
  where the narrowest part, the throat, is simply a minimum of the spherical radius $r$. 
  A majority of known exact \wh\ solutions both in general relativity (GR) and alternative theories 
  of gravity (e.g., \cite{almaz1, almaz2, EGB, b-kim03, HOG-13}) are \sph.
  
  In studies of macroscopic phenomena or possible artificial constructions, if our interest is in 
  describing (potentially) realistic and manageable \whs, there is a good reason to adhere to GR 
  since it is this theory that is well verified by experiment at the macroscopic level and even serves
  as a tool in a number of engineering applications such as, for instance, global positioning systems.
  Then, the existence of traversable Lo\-rentz\-ian \whs\ as solutions to the Einstein equations 
  requires some kind of ``exotic matter'', i.e., matter that violates the null energy condition 
  (NEC) \cite{thorne, hoh-vis}, which is in turn a part of the weak energy condition (WEC) whose 
  physical meaning is that the energy density is nonnegative in any reference frame. 

  Many \sph\ \wh\ solutions in GR were first obtained with scalar field sources, beginning with a 
  massless phantom scalar field \cite{br73, ellis}, and were later extended to include scalar field 
  potentials as well as electromagnetic and other fields, see, e.g., 
\cite{br73, BBS12, ju-dil, anab12, BKor15, k17-book, balak10} and references therein. 
  Meanwhile, there appeared quite numerous examples of \wh\ solutions with fluid sources 
  with various equations of state or with unspecified matter whose stress-energy tensor (SET) 
  components are formally called the density $\rho$ and pressure $p$, and the latter may be 
  different in different directions --- see, e.g, 
\cite{b-kim03, sush-05, lobo-05, rah-06, lobo-12, lobo-chapl, catal, kuh17}.
  An evident difficulty in this interpretation is that in such fluids in many cases (such as, 
  for instance, $p = w\rho$ with $w < -1$) the velocity of sound calculated as $(dp/d\rho)^{1/2}$ 
  turns out to be imaginary, which leads to a hydrodynamic instability with exponentially growing 
  perturbations. A way out is to suppose that what we call a ``fluid'' actually consists of some 
  fields with quite different perturbation dynamics.  

  These difficulties are absent (though others may appear) if, instead of standard GR, we adhere to the brane
  world concept. Let us recall that in brane world theories (see, e.g., the reviews \cite{BW1,BW2,BW3} 
  and references therein) the observable 4D world is a kind of domain wall in five or more dimensions
  of large or even infinite size. The standard-model fields are confined on the brane while gravity
  propagates in the surrounding bulk. The gravitational field on the brane itself can be described, 
  at least in a large class of models related to Randall and Sundrum's second model \cite{RS2}
  (RS2), by modified 4D Einstein equations \cite{SMS, BW3}, where, in addition to $T\mn$ ---
  the SET of the 4D matter and a cosmological term $\Lambda_4 g\mn$,  there is also a tensor quadratic 
  in $T\mn$, a contribution from bulk matter (if any), and a geometric term $E\mn$ representing a 
  projection of the ``electric'' part of the 5D Weyl tensor onto the brane. In vacuum, when both 4D and 
  5D matter is absent, these equations  take the form $G\mn + \Lambda_4 g\mn = - E\mn$, where $G\mn$ 
  is the 4D Einstein tensor. The tensor $E\mn$, connecting gravity on the brane with the bulk geometry, 
  sometimes called the tidal SET, is traceless by construction. Due to its geometric origin, it is not subject 
  to requirements like energy conditions or hydrodynamic stability. The form of $E\mn$, apart from its 
  zero trace, is virtually arbitrary (as guaranteed by the known embedding theorems), so, if 
  $\Lambda_4 = 0$ (a reasonable assumption for describing local objects), $R=0$ is the only unambiguous
  consequence of the equations $G\mn = - E\mn$.

  A class of \asflat\ \wh\ solutions to the equation $R=0$ was obtained in \cite{b-kim03}, and 
  it may be on equal grounds interpreted as describing vacuum gravitational fields in a brane world, 
  existing due to a``tidal'' influence from the fifth dimension, or GR \whs\ supported by anisotropic fluids 
  with a traceless SET. Other examples of solutions with $R =0$ appeared in \cite{dad02, agn02}, their 
  non-vacuum extensions in \cite{bbh-03, lobo-07, molina12}, in particular, with $\Lambda_4\ne 0$ 
  in \cite{molina12}.

  A common feature of these and many other \wh\ solutions is that they were obtained using the 
  spherical radius $r$ that has a minimum on the throat, as the radial coordinate, and, as a result, 
  the solutions cover only one half of the \wh\ space-time, the other half being its copy. Thus such 
  \whs\ are by construction $\Z_2$-symmetric with respect to their throats. Obtaining asymmetric \whs\ 
  using this coordinate is possible but technically rather inconvenient, see details in \cite{b-kim03},
  and there are very few examples of such \wh\ solutions \cite{b-kim03, dad02, agn02, balak10}.

  Another common feature of solutions with fluid sources is that the fluid is either anisotropic 
  or, if isotropic, occupies a finite volume bounded by a junction surface, with a vacuum metric 
  outside it. In most of the cases the junction surface is a thin shell with its own surface density
  and pressure. While such inclusions are often convenient for model construction, they look 
  rather artificial, and it is worthwhile to consider matter distributions either directly adjoint
  to vacuum through a usual boundary, like a stellar surface, or gradually decaying and approaching 
  vacuum at infinity. 
  
  In all cases, a vacuum geometry at infinity may be either flat, which is reasonable for describing
  a \wh\ in the modern Universe, or (anti-) de Sitter ((A)dS), which is more suitable for possible
  \whs\ in an inflationary universe or those related to vacuum bubbles  (see, e.g., the recent 
  paper \cite{seno-17} and references therein). One might claim that a static configuration with
  a dS asymptotic cannot be a \wh\ because of inevitable existence of horizons; such horizons,
  however, are of cosmological rather than \bh\ nature, so it makes sense to widen the \wh\ notion
  by admitting them. In an inflationary universe, such \whs, connecting otherwise distant and causally
  disconnected regions of dS space, may in principle contribute to solving the horizon problem in
  cosmology, diminishing the necessary number of e-folds. Another application can be 
  the modern accelerated Universe (or two such universes as possible locations of \wh\ mouths) 
  \cite{lemos03, su-kim04, su-zhang07, balak10, molina11, anab12, moke13}. 

  In this paper we will show that \ssph, \asflat\ or AdS \wh\ geometries do not exist in GR with 
  any kind of matter with isotropic pressure as a source (Section 2). On the contrary, asymptotically 
  dS configurations are allowed, and we will construct examples of such \wh\ solutions using the 
  so-called quasiglobal coordinate $x$ \cite{BR-book, vac1} such that $g_{tt}g_{xx} = -1$. 
  The examples include both $\Z_2$-symmetric and asymmetric \whs\ (Section 3). In Section 4 
  we present examples of symmetric and asymmetric \asflat\ \whs\ under the assumption $R =0$. 
  We note that although the curvature coordinate $r$ is often favorable for obtaining analytic 
  solutions (in \cite{b-kim03} and \cite{bbh-03} different \wh\ and \bh\ solutions were obtained
  in an algorithmic form), the coordinate $x$ is more natural and transparent when dealing with 
  \bh\ and \wh\ geometries, even though our solutions in terms of $x$ are only numerical.  
  As a by-product, in our study there emerge a number of regular, twice \asflat\ \bh\ solutions 
  with up to four Killing horizons. In Section 5 we consider the traversability properties of the twice
  \asflat\ \whs\ and calculate the light deflection angles in their geometries, leading to gravitational 
  lensing.  Section 6 contains some concluding remarks. Finally, in the Appendix we discuss 
  an interesting property of the integral curves of our equations for the redshift function 
  $A(x) = g_{tt}$: all curves beginning at the throat $x=0$ with the same slope $A'$ but  
  different values of $A(0)$, intersect at certain values of $x$. It turns out to be a manifestation of 
  a general property of linear differential equations.

% ===================================
\section{\nhq Basic equations. No-go theorem}
% -----------------------------------------------
\subsection {General relations. The necessity of NEC violation}

  Let us begin with the general \ssph\ metric which can be written in the form\footnote
      {Our conventions are: the metric signature $(+\ -\ -\ -)$, the curvature tensor
        $R^{\sigma}{}_{\mu\rho\nu} = \d_\nu\Gamma^{\sigma}_{\mu\rho}-\ldots,\
        R\mn = R^{\sigma}{}_{\mu\sigma\nu}$, so that the Ricci scalar
        $R > 0$ for de Sitter space-time and the matter-dominated
        cosmological epoch; the sign of $T\mN$ such that $T^0_0$ is the energy
        density, and the system of units $8\pi G = c = 1$.}
\beq                                                           \label{ds}
        ds^2 = \e^{2\gamma(u)}dt^2 - \e^{2\alpha(u)}du^2 - \e^{2\beta(u)} d\Omega^2,
\eeq
  where $u$ is an arbitrary radial coordinate and  
  $d\Omega^2 = d\theta^2 + \sin^2 \theta d\varphi^2$ is the linear element on a unit 
  sphere.\footnote
	{We use different letters for different radial coordinates: $u$ is a general notation,
	  without a specific ``gauge'' condition, $x$ is a quasiglobal coordinate, 
           such that $\alpha = -\gamma$ in \rf{ds}, and $l$ is the Gaussian, or proper 
           radial distance coordinate, such that $\alpha \equiv 0$ in \rf{ds}.}

  Then the Ricci tensor has the following nonzero components: 
\bearr \nhq
       R^t_t =                                         \label{R00}
               -\e^{-2\alpha}[\gamma'' +\gamma'(\gamma'-\alpha'+2\beta')],
\yyy \nhq
      R^u_u = \!\!               \                          \label{R11}
     {-}\e^{-2\alpha}[\gamma''+2\beta'' +\gamma'{}^2+2\beta'{}^2
%\nnn \inch
            -\alpha'(\gamma' \!+\! 2\beta')],
\yyy  \nhq
     R^\theta_\theta = R^\varphi_\varphi
		= \e^{-2\beta}                                    \label{R22}
               -\e^{-2\alpha}[\beta''+\beta'(\gamma'-\alpha'+2\beta')] ,
\ear
  where the prime stands for $d/du$. The Einstein equations can be written in two equivalent forms
\bearr                                                                   \label{EE}
    G\mN \equiv R\mN - \half \delta\mN R = - T\mN, \qquad {\rm or}
\nnn 
    R\mN = - (T\mN - \half \delta\mN T^\alpha_\alpha), 
\ear
  where $T\mN$ is the stress-energy tensor (SET) of matter.
  The most general SET compatible with the geometry \rf{ds} has the form 
\beq
	T\mN = \diag (\rho,\ -p_r,\ -p_T,\ -p_T),
\eeq
  where $\rho$ is the energy density, $p_r$ is the radial pressure, and $p_T$ is the 
  tangential pressure. These SET components may contain contributions of one or several
  physical fields of different spins and masses but can also be considered as hydrodynamic
  quantities, characterizing the density and pressures of one or several fluids. In this paper
  we consider $T\mN$ as the SET either in a general form or as that of a single fluid, which 
  is in general anisotropic ($p_r \not\equiv p_T$).
    
  Let us illustrate the necessity of exotic matter for \wh\ existence in space-times 
  with the metric \rf{ds} taken as an example \cite{thorne, BR-book}. Choosing the so-called 
  quasiglobal coordinate $u = x$ under the condition $\alpha+\gamma =0$ and denoting
  $\e^{2\gamma} = \e^{-2\alpha} = A(x)$, $\e^\beta = r(x)$, we rewrite the metric as
\beq          \label{ds-q}
           ds^2 = A(x) dt^2 - \frac{dx^2}{A(x)} - r^2 (x) d\Omega^2.  
\eeq
  A (traversable) wormhole geometry implies, by definition, that the function $r(x)$ has a regular 
  minimum (say, at $x=x_0$), called a throat, and reaches values much larger than $r(x_0)$ on 
  both sides of the throat. It is also usually required that $A(x) > 0$ in the whole range of $x$, which 
  excludes horizons that characterize \bh\ rather than \wh\ geometries. It may happen, however, that 
  there is a horizon far away from the throat, for example, if the space-time is \asdS. If a \wh\ connects 
  two de Sitter universes or regions of a single de Sitter universe, it will have horizons on each side
  of the throat, but the latter lies in the region where $A(x) > 0$. 
  
  The difference of the  ${t\choose t}$ and ${x\choose x}$ components of the Einstein 
  equations reads 
\beq          \label{01q}
      2A\, r''/r = - (T^t_t - T^x_x) \equiv - (\rho + p_r),        
\eeq
  On the other hand, at a throat as a minimum of $r(x)$ we have
\beq
     r  > 0, \qquad       r' =0, \qquad         r''> 0.                   \label{min}
\eeq
  (In special cases where $r''=0$ at the minimum, it always holds $r''> 0$ in its 
   neighborhood.) Then from \rf{01q} it immediately follows  $\rho + p_r < 0$. This inequality 
  does indeed look exotic, but to see an exact result, let us recall that the NEC requires
  $T\mN k^\mu k_\nu \geq 0$, where $k^\mu$ is any null vector, $k^\mu k_\mu = 0$. Choosing 
  $k^\mu = (1/\sqrt{A}, \sqrt{A}, 0, 0)$, we see that  $T\mN k^\mu k_\nu = \rho+p_r$.
  Thus the inequality $\rho + p_r < 0$ does indeed violate the NEC. 

% -------------------------------------------------------------
\subsection {No-go theorem} 

  There are three different nontrivial components in the Einstein equations for 
  the metric \rf{ds-q}, written using the quasiglobal gauge $\alpha + \gamma=0$:
\bearr            \label{E00}
      G^t_t = \frac{1}{r^2}[-1 +A(2rr'' +r'^2) +A'rr'] = - T^t_t,
\yyy                                                                \label{E11}
      G^x_x = \frac{1}{r^2}[-1 + A' rr' + Ar'^2] = - T^x_x,
\yyy							        \label{E22}
      G^\theta_\theta =  G^\phi_\phi = \frac{1}{2r}[2A r'' + rA'' + 2A'r'] = p_T,
\ear
  where the prime again denotes $d/dx$, and \rf{E11} is the constraint equation, free from 
  second-order derivatives. Note that $T^t_t$ is the energy density $\rho$ 
  and $T^x_x = - p_r$ in a static (R-) region, where $A > 0$, while beyond a horizon 
  (if any), in a T-region where $A(x) < 0$, where the metric describes a \KS\ cosmology,
  $T^x_x$ is the energy density, and $- T^t_t$ is the pressure in the (spatial) $t$ direction.   

  It is of interest whether or not \wh\ solutions can be obtained with a source in the
  form of an isotropic (Pascal) fluid, such that $p_r = p_T$. Let us show that the answer is
  negative if we require $A > 0$ in the whole space.

  If $p_r = p_T$, it follows $G^x_x =G^\theta_\theta$, and the difference of \eqs \rf{E11}
  and \rf{E22} gives
\beq                                                   \label{iso1}
	r^2 A'' + 2A r r'' -2A r'^2 + 2 =0.
\eeq 
  The substitution $A(x) = D(x)/r^2(x)$ converts it to
\beq                                                   \label{iso2}
         D'' - \frac{4 D' r'}{r} + \frac{4 D r'^2}{r^2} + 2 =0.   
\eeq
  A possible minimum of $D(x)$ at some $x=x_0$ requires $D'=0$ and $D'' \geq 0$. Meanwhile,
  if $D'= 0$,  \eqn{iso2} gives $D'' \leq -2$, so that it is necessarily a maximum. 

  However, an \asflat\ traversable \wh\ requires $r\to \infty$ and $A \to 1$ as $x \to \pm\infty$,
  in an \asAdS\ \wh\ it must be $A \sim r^2$ at large $r$, etc. In all such cases 
  $D(x) \to \infty$ on both sides far from the throat, hence it should have a minimum, which, 
  as we have seen, is impossible. We thus have the following theorem:

{\sl 
  A \ssph\ traversable wormhole with $r\to\infty$ and $A(x) r^2(x)\to \infty$  on both sides of 
  the throat cannot be supported by any isotropic matter source with $p_r = p_T$.    
 }

  This excludes, in particular, twice \asflat\ and twice asymptotically AdS \whs\  as well as those 
  \asflat\  on one end and AdS on the other. What is not excluded, is that one or both asymptotic 
  regions are de Sitter: in this case, $r \to \infty$ but $A \sim -r^2$ at large $r$, and it is not 
  necessary to have a minimum of $D(x)$.

  The theorem has been proved using a specific coordinate condition but has an invariant meaning 
  since the quantities $A= g_{tt}$ and $r^2 = g_{\theta\theta}$ are insensitive to the choice of the 
  radial coordinate. The existing \wh\ solutions where an isotropic fluid occupies a finite region of 
  space also do not contradict the theorem since they inevitably require ``heavy'' thin shells on 
  the boundary between the fluid and vacuum regions \cite{sush-05, lobo-05}, and these shells are 
  highly anisotropic in the sense that a tangential pressure is nonzero while the radial one is not 
  defined (the radial direction is orthogonal to the shell).

% =============================
\section{Wormholes supported by\\ isotropic matter}
% --------------------------------------------------
\subsection{Symmetric dS--dS wormholes}
%--------------------------------------------
  
  The opportunity of obtaining \whs\ with two de Sitter asymptotics (dS--dS \whs\ for short)
  supported by isotropic matter, allowed by the above no-go theorem, is of interest, in particular, 
  because such \whs\ may exist in an inflationary universe and provide causal connections between 
  otherwise distant regions. We will construct examples of such solutions in this subsection.
 
  Another opportunity of interest is a \wh\ connecting de Sitter regions with different values 
  of the cosmological constant, which may be interpreted as bubbles of true and false vacua.
  In such cases a \wh\ can either play the role of a thick domain wall, or, on the contrary,
  directly connect regions separated by a domain wall. Such examples will be discussed
  in the next subsection. 

  Since there is no clear reason to assume any particular equation of state, we will instead specify 
  the metric function $r(x)$ having a regular minimum at $x=0$ (the throat) and compatible 
  with a de Sitter behavior of the metric at large $|x|$:
\beq                                                    \label{r_x}
		r(x) = \sqrt{a^2 + x^2}, \qquad a = \const > 0.
\eeq
  For a numerical study, we put $a = 1$; remaining arbitrary, the parameter $a$ will then play 
  the role of a length unit. Furthermore, assuming that the matter source is isotropic, $p_r = p_T$, 
  we can use \eqn{iso1} for finding $A(x)$; after solving it, the metric will be known completely.

  With \rf{r_x} and $a=1$, \eqn{iso1} takes the form
\beq                                                   \label{iso3}
	      (1+x^2)^2 A'' + 2 (1-x^2) A + 2 (1+x^2) =0. 
\eeq
  It is hard to solve this equation analytically, but its asymptotic form at large $|x|$, that is,
  $x^2 A'' -2A + 2 = 0$, is easily integrated giving $A(x) = 1 + c_1 x^2 + c_2/x$ with 
  $c_{1,2} = \const$. Solutions with $c_1 \geq 0$, corresponding to flat or AdS asymptotics, 
  are excluded by the above no-go theorem, so the only possible asymptotic form of the metric 
  is de Sitter, with $c_1 < 0$.

  Examples of numerical solutions to \eqn{iso3} under the initial conditions $A(0) = A_0$,
  $A'(0) =0$ are shown in Fig.\,1 for $A_0 = 1, 3, 6, 10, 15$. It is of interest that all curves 
  intersect at two symmetric points: $x \approx \pm 1.4109$, $A(x) \approx -1.4953$.
% ------------------------------------------------ fig 1
\begin{figure}[ht]
\centering
\includegraphics[width=6.5cm]{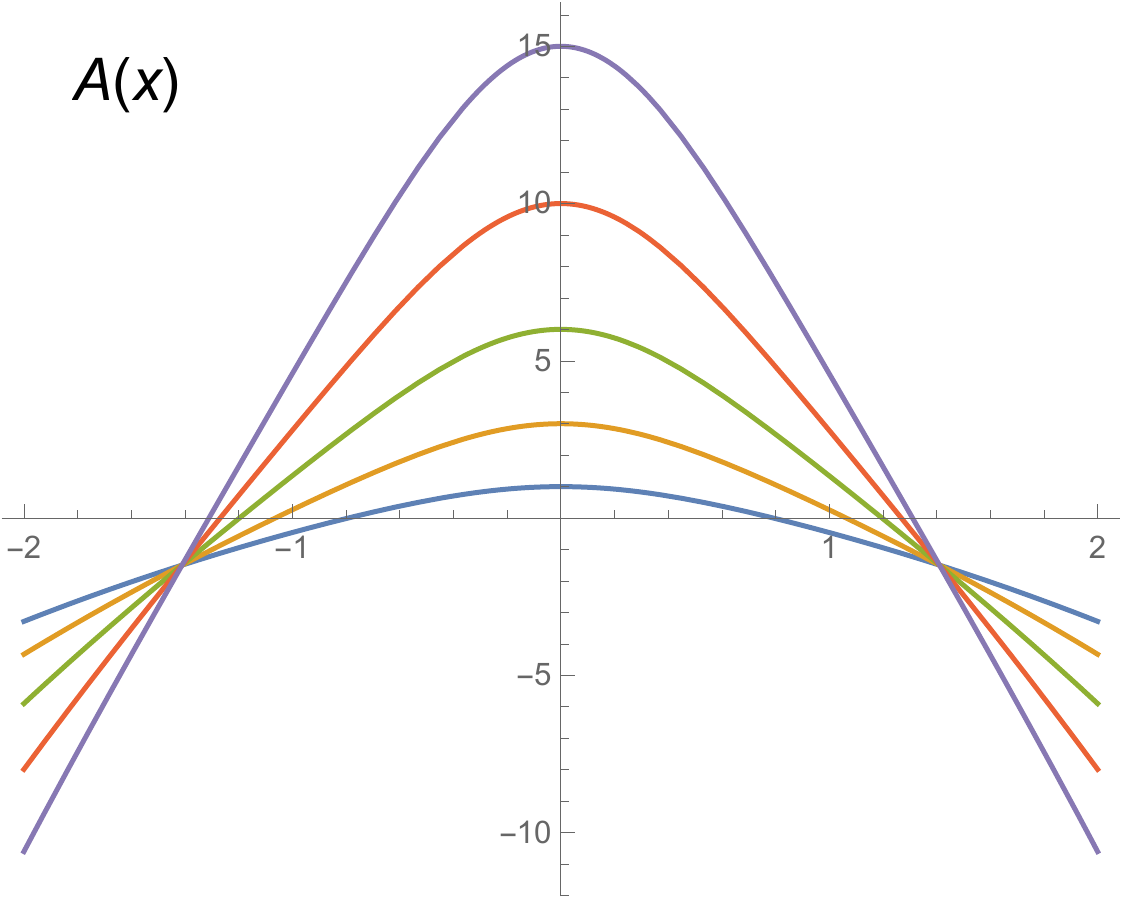}
\caption{\small
        Solutions $A(x)$ of \eqn{iso3} for a symmetric dS--dS \wh, corresponding to 
        $A(0) = 1, 3, 6, 10, 15$ (bottom-up along the ordinate axis) and $A'(0) =0$. }
\end{figure}
 % ------------------------------------------------ fig 2
\begin{figure}[ht]
\centering
\includegraphics[width=7cm]{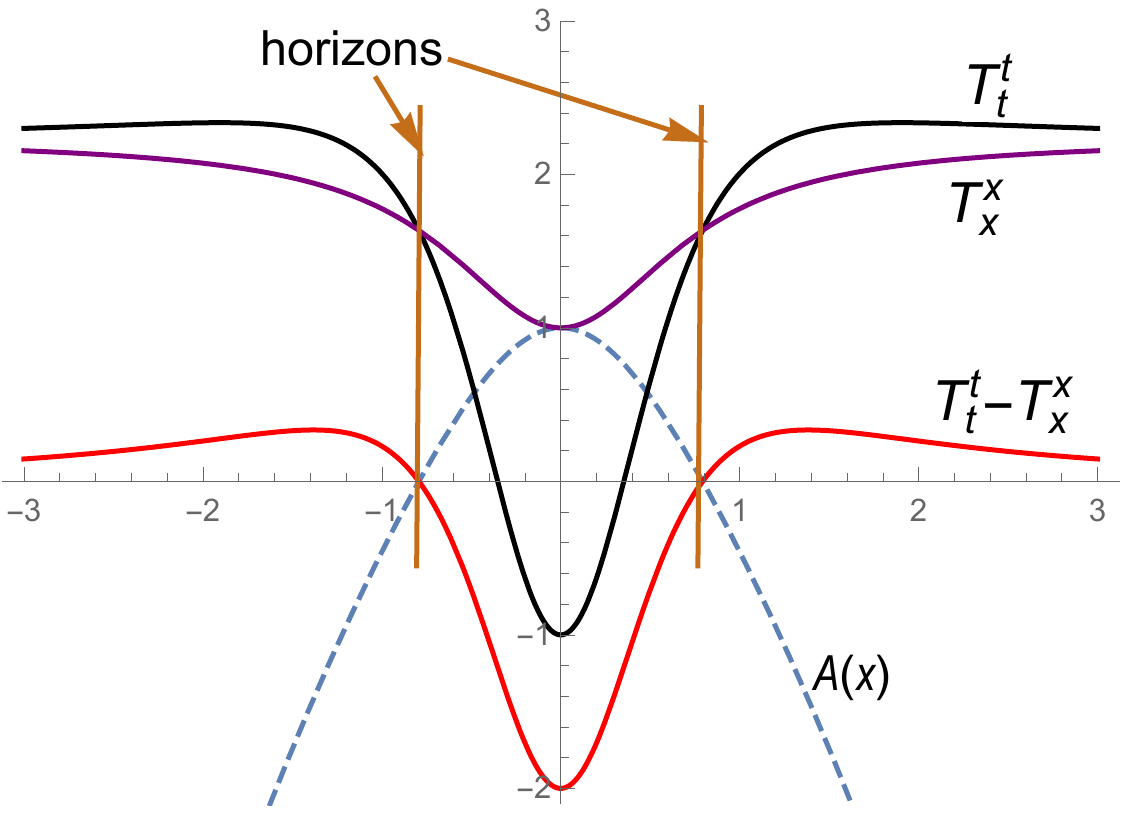}
\caption{\small  
         The metric function $A(x)$ and the SET components for a symmetric dS--dS \wh\ 
         described by the solution of \eqn{iso3} with $A(0) = 1$ and $A'(0) =0$. }
\end{figure}
% --------------------------------------------------
  The behavior of SET components in the \wh\ model with  $A(0) = 1$ and 
  $A'(0) =0$, found from \eqs \rf{E00} and \rf{E11}, is shown in Fig.\,2.
  The \wh\ is $\Z_2$-symmetric relative to its throat ($x = 0$), so all these functions are even.
  The values of the effective cosmological constant $\Lambda$ at large $|x|$ correspond to 
  $A(x) \approx -\Lambda x^2/3$, thus it is clear from Fig.\,1 that $\Lambda$ is the same at the 
  two infinities but different for different values of $A(0)$. Its precise value is in each case 
  determined as the common limit of $T^t_t$ and $T^x_x$, as exemplified by Fig.\,2.  

  Note that the usual relations $T^t_t = \rho$ and $T^x_x = -p$ hold only in the static region,
  where $A(x) > 0$. In a region where $ A(x) < 0$ (T-region), the coordinate $t$ is spatial, 
  hence $-T^t_t  = p_t$  is the pressure along the $t$ direction, while the density is  
  $\rho = T^x_x$ since $x$ is now a temporal coordinate; however, the condition $r'' > 0$ in 
  \eqn{01q} (which is the same for any sign of $A$) leads to $\rho + p_t < 0$, again
  violating the NEC. That $T^t_t$ and $T^x_x$ tend to the same constant value at large $x$  
  agrees with the de Sitter asymptotic behavior of the metric since the SET structure approaches
  that of a cosmological term, $T\mN = \Lambda \delta\mN$. One can also notice that this
  structure takes place on the horizon. Furthermore,  in the whole space-time we have 
  $T^x_x = T^\theta_\theta = -p_T$,  but the fluid is anisotropic in the T-region since 
  $p_t = -T^t_t \ne p_T$. 

  Fig.\,3 shows the Carter-Penrose diagram of dS-dS \whs, the same as presented in 
  \cite{lemos03, moke13}, and potentially infinite both to the left and to the right: however, 
  mutually isometric surfaces may be identified, such as, e.g., those depicted by lines AA' and BB'
  in Fig.\,3, which means that the \wh\ connects regions of the same de Sitter universe.    

 % ------------------------------------------------ fig 3
\begin{figure}[ht]
\centering
\includegraphics[width=7cm]{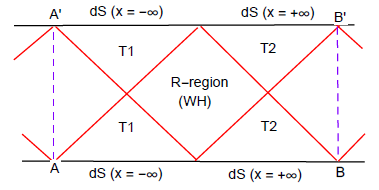}
\caption{\small  
        Carter-Penrose diagram showing the global structure of dS-dS \whs. 
        The regions labeled T1 correspond to the range $x < x_-$, those labeled T2 
        to the range $x > x_+$, where $x_- < x_+$ are the horizons (and $x_- = - x_+$ 
        in a symmetric solution only). 
}
\end{figure}
% --------------------------------------------------
 
% --------------------------------------------
\subsection{Asymmetric configurations: dS--dS wormholes and black universes}
%--------------------------------------------
 
 % ------------------------------------------------ fig 4
\begin{figure}[ht]
\centering
\includegraphics[width=7cm]{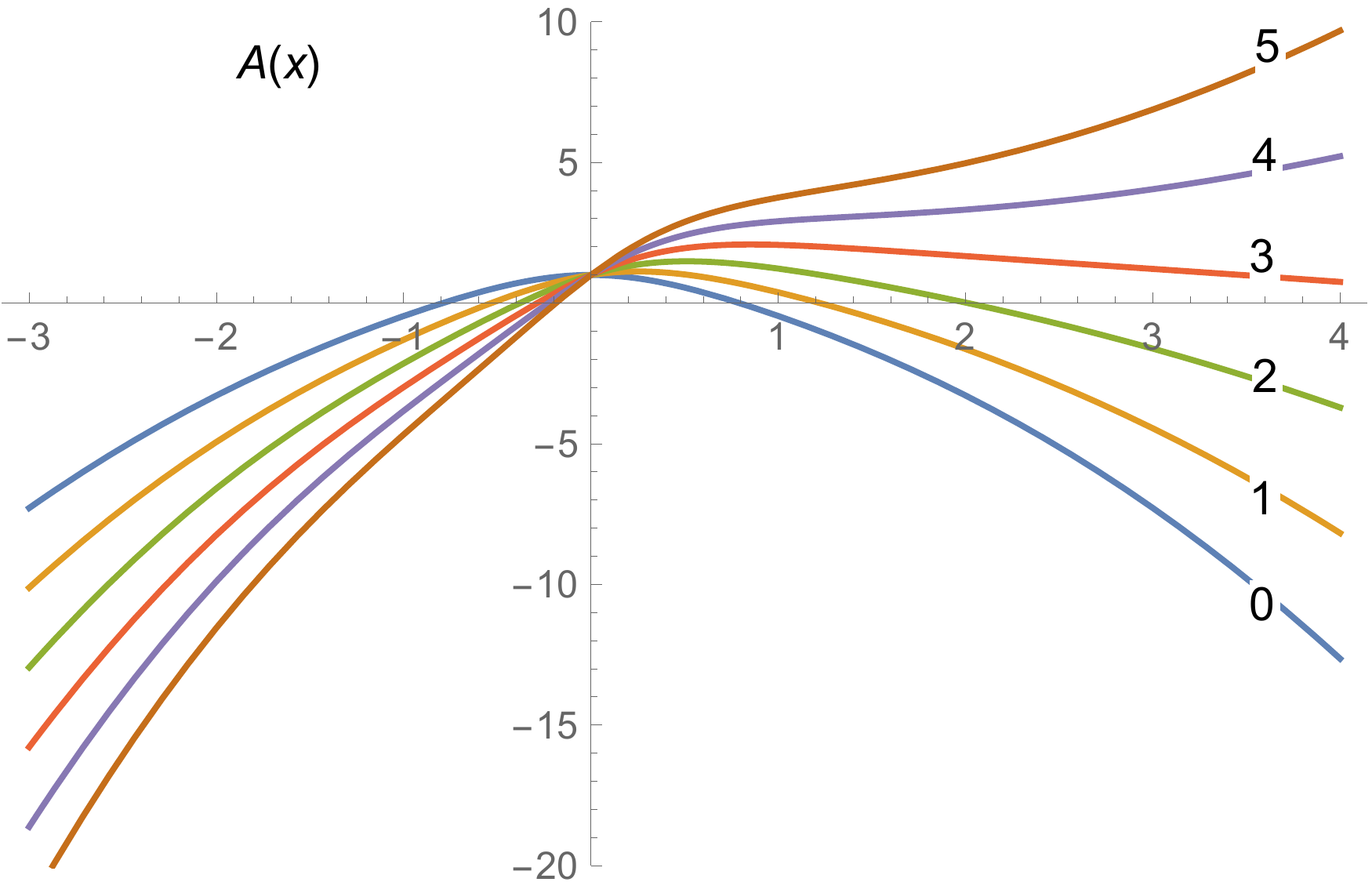}
\caption{\small  
         The function $A(x)$ for $A(0) = 1$ and $A'(0) = 0, 1, 2, 3, 4, 5$ (the latter values are 
	written on the corresponding curves).}
\end{figure}
 % ------------------------------------------------ fig 5
\begin{figure}[ht]
\centering
\includegraphics[width=6.5cm]{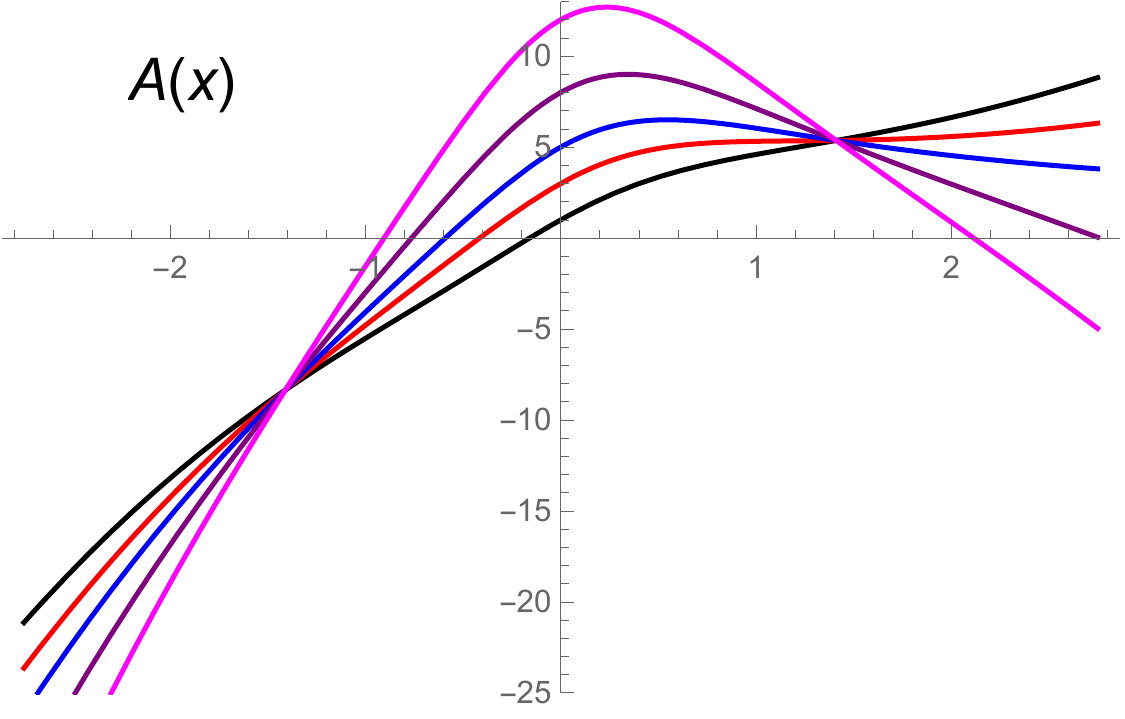}
\caption{\small  
         The function $A(x)$ for the same slope at the throat, $A'(0) = 6$ and $A(0) = 1, 3, 5, 8, 12$ 
         (bottom-up along the ordinate axis and conversely at large $|x|$)} 
\end{figure}
% ------------------------------------------------ fig 6
\begin{figure}[h]
\centering
\includegraphics[width=6.5cm]{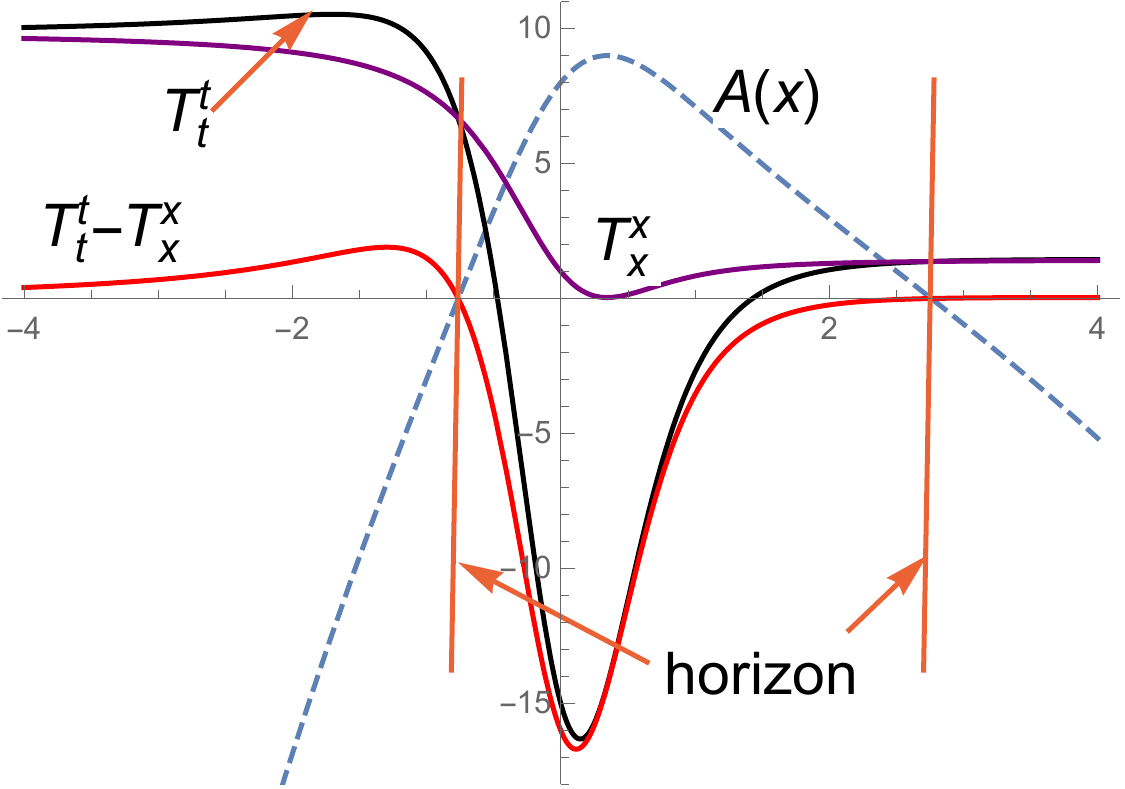}
\caption{\small  
         The function $A(x)$ and the SET components for the solution with $A(0) = 8$ and 
           $A'(0) = 6$, having two de Sitter asymptotics.with different curvature values.}
\end{figure}
% ------------------------------------------------ fig 7
\begin{figure}[h!]
\centering
\includegraphics[width=6.5cm]{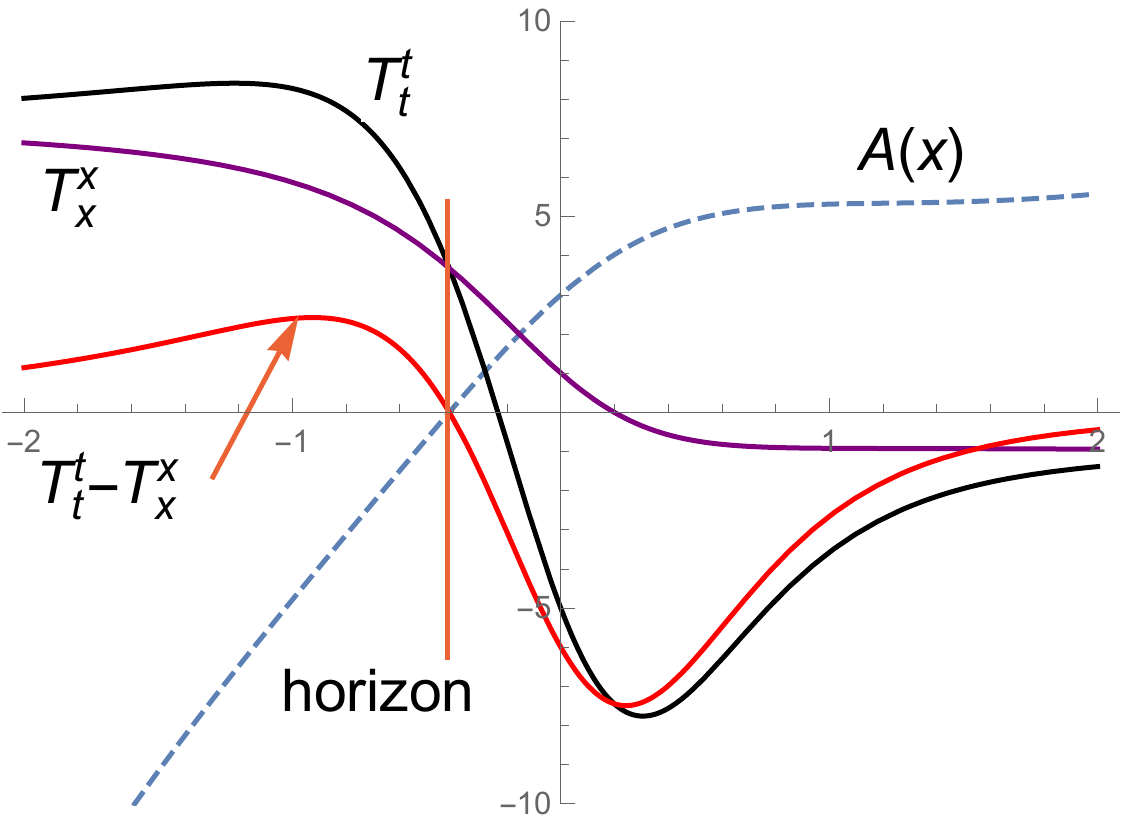}
\caption{\small  
         The same as in Fig.\,6 but $A(0) =3$ and $A'(0) =6$, with AdS behavior at large positive $x$.} 
\end{figure}
% --------------------------------------------------  

 Solutions to the same equation \rf{iso3} with $A'(0) \ne 0$ can also be obtained 
  numerically. Some examples of such solutions are shown in Fig.\,4 for $A(0) = 0$  
  and $A'(0) = 0, 1, 2, 3, 4, 5$. These plots show that at small values of $A'(0)$ we obtain
  dS-dS \whs\ with different values of $\Lambda$ at large positive and negative $x$. Such \whs\
  might connect space-time regions with different vacuum energy densities, for instant, a bubble of
  false vacuum with a region of true vacuum. The global structure diagram for all such space-times
  is the same as for symmetric dS-dS \whs.

  Larger values of $A'(0)$ lead to configurations with a single horizon and an AdS asymptotic behavior 
  as $x\to\infty$. There is an intermediate case with asymptotic flatness at large positive $x$,
  as is proved by the existence of solutions to \eqn{iso3} under the condition, e.g., $A(+\infty) = 1$. 
  All such configurations have the structure of black universes \cite{bu1, bu2, BBS12, BKor15}, 
  i.e., \bhs\ in which beyond the horizon there is, instead of a singularity, an expanding universe 
  tending to a de Sitter behavior at late times. 

  Figure 5 shows how the behavior of $A(x)$ changes if one keeps invariable the derivative $A'(0)$
  and changes $A(0)$. Again there are different asymptotic behaviors as $x\to +\infty$ depending 
  on the value of $A(0)$. But again, just as in Fig.\,1, all plots of $A(x)$ intersect at two points.  
  More than that, these intersections occur at the same values of $x \approx \pm 1.4109$ as 
  it happened for symmetric models, though the corresponding values of $A(x)$ are certainly 
  different. All this cannot happen by chance, and indeed, one can prove that it is a manifestation 
  of a general property of second-order linear differential equations, see the Appendix. 

  The SET components behave accordingly. Figs. 6 and 7 show their properties for two different 
  cases, one for a dS-dS \wh\ with different dS curvatures at two ends (Fig.\,6), and the other 
  where the right end is AdS. The corresponding values of the effective cosmological constant 
  are the same as those of the SET components since at large $|x|$ all of them coincide (recall
  that we are here dealing with isotropic matter, hence 
  $T^x_x = T^\theta_\theta = T^\varphi_\varphi$).   

% =====================================================
\section {Asymptotically flat (M--M) wormholes and regular \bhs}

% ------------------------------------------------ fig 8
\begin{figure*}[t]
\centering
\includegraphics[width=6cm]{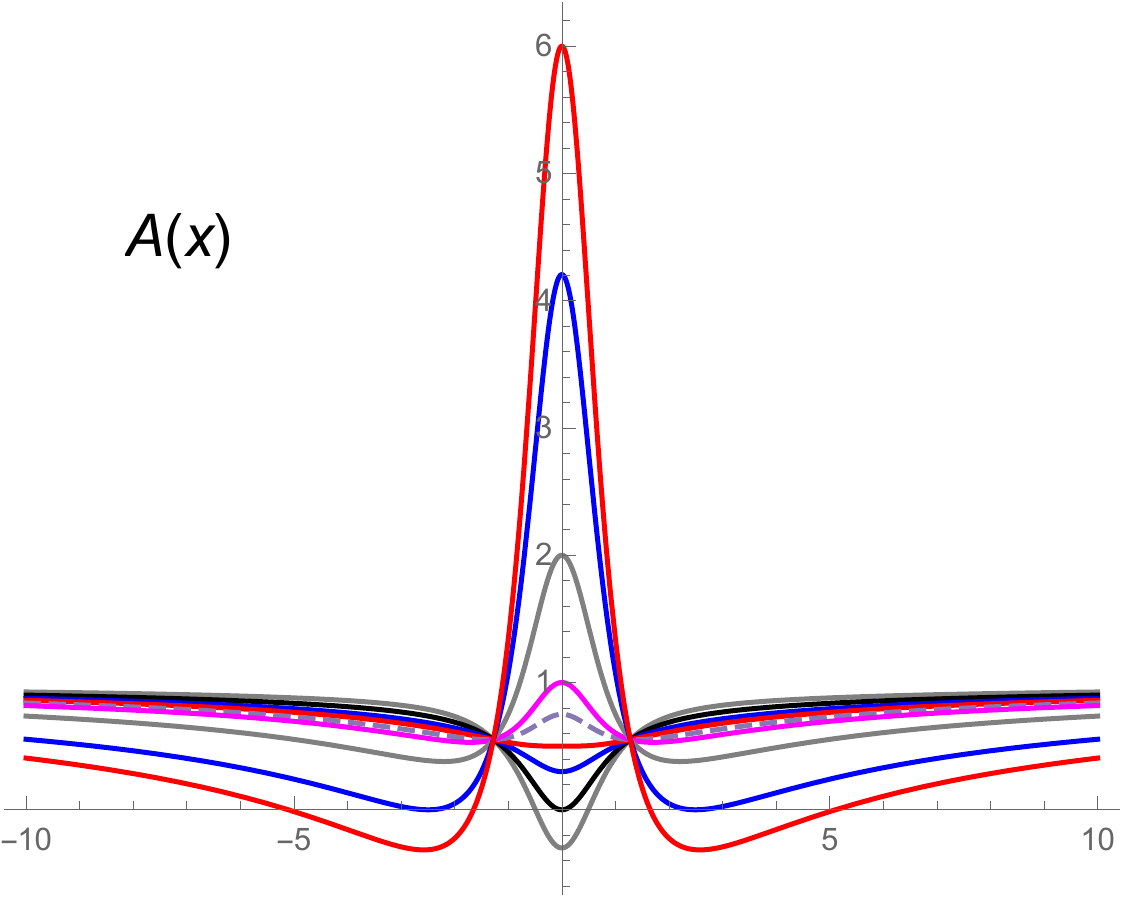} \
\includegraphics[width=5.5cm]{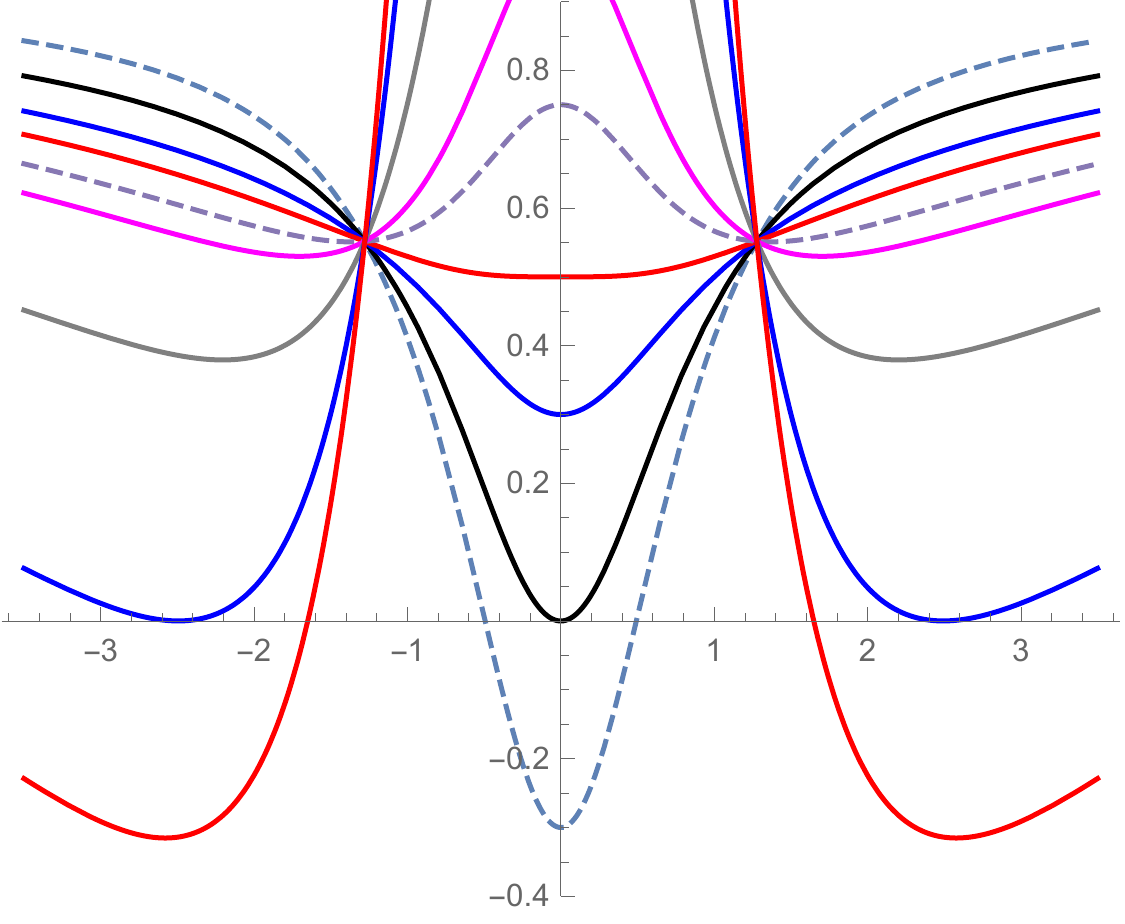} \
\includegraphics[width=5.5cm]{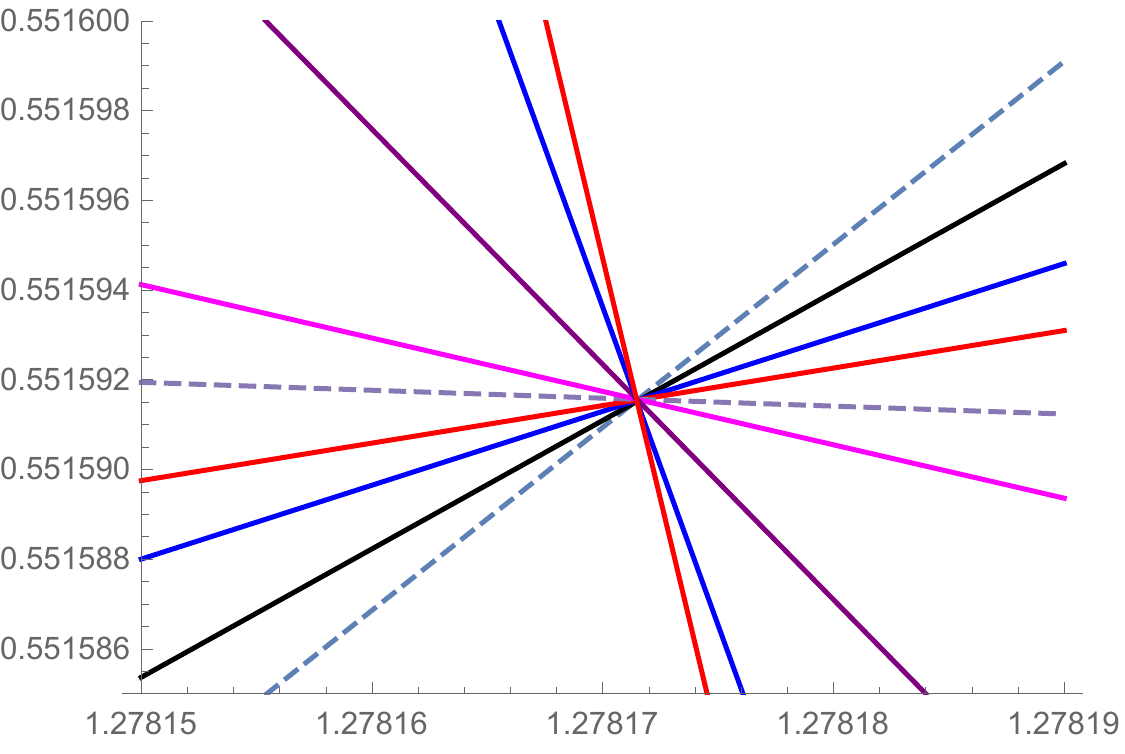}
\caption{\small  
         Symmetric solutions to \eqn{eq-R0} with $A'(0)=0$ and 
         $A(0) = -0.3,\ 0,\ 0.3, 0.5,\ 0.75,\ 1,\ 2,\ 4.205,\ 6$
         (bottom-up along the ordinate axis, and conversely at large $|x|$). 
         Left --- a general picture, middle ---  its part of interest enlarged.
         right --- a neighborhood of the plots' intersection point ``under a microscope''. 
}      \label{Aa-sym}
\end{figure*}
% ------------------------------------------------ 

  Now let us abandon the source isotropy assumption and try to obtain some new models of 
  twice \asflat\ \wh\ geometries. Since, as before, there is no clear reason to assume a particular
  form of the equations of state (which are now different for $p_r$ and $p_T$), let us, instead, 
  again choose the function $r(x)$ in the form \rf{r_x}. In addition, let us assume a zero scalar
  curvature $R$ throughout the space. In this way we not only replace postulating another 
  equation of state of the source matter, but also make it possible to interpret the results as
  vacuum solutions in an RS2-like brane world, somewhat similar to those found in \cite{b-kim03},
  where some examples of $\Z_2$-symmetric \wh\ solutions were obtained in an analytic form 
  using the spherical radius $r$ as a coordinate. Now we will use the coordinate $x$ which is 
  better for finding $Z_2$-asymmetric solutions, although these solutions will be only numerical.  

  For the metric \rf{ds-q} we have
\beq           \label{R-q}
	R = \frac{2}{r^2} - A'' - 4A'\frac{r'}{r} -4A\frac{r''}{r}- 2A \frac{r'^2}{r^2}.
\eeq       
  Therefore, under the assumption \rf{r_x} for $r(x)$, with, as before $a=1$, the equation $R=0$ 
  takes the form
\beq                                    \label{eq-R0}
	  A'' + \frac{4x}{1+x^2} A' + \frac{2(2+x^2)}{(1+x^2)^2} A = \frac{2}{1+x^2}.
\eeq
  At large $|x|$ the asymptotic form of this equation has the general solution 
  $A = 1 + C_1/x + C_2/x^2$, $C_{1,2}= \const$, which evidently corresponds to asymptotic 
  flatness with a \Scz-like metric. So, let us solve \eqn{eq-R0} under the initial conditions specified 
  at $x=0$: $A(0)$ and $A'(0)$, so that $A'(0)=0$ should lead to an even function $A(x)$, hence a
   symmetric solution, and $A'(0) \ne 0$ to an asymmetric one.

  Examples of symmetric solutions to \eqn{eq-R0} with different $A(0)$ and $A'(0) =0$ are 
  plotted in Fig.\,\ref{Aa-sym}. It is observed that at $0 < A(0) = 0.5$ we have wormholes 
  with a minimum of $A(x)$ at the throat $x=0$, which is thus attracting for test particles. At 
  larger $A(0)$ there appear two minima of $A(x)$ around the throat, acting as potential wells for 
  test particles. A further increase of $A(0)$ makes these minima first equal to zero (at $A(0) 
  \approx 4.205$) and then negative at still larger $A(0)$. We thus obtain regular black holes 
  with either two double horizons or four simple horizons. A phenomenon of interest, as in Section 3, 
  is that $A(\pm 1.278172)\approx 0.551892$ independently of $A(0)$, which is observed as the 
  existence of two intersection points of all plots in Fig.\,\ref{Aa-sym}. On the other hand, the 
  initial values $A(0) = 0$ and $A(0) < 0$ lead to solutions with one double or two simple horizons,
  respectively, similar to those found in \cite{bbh-03}. Thus we are again dealing with regular \bhs\ 
  instead of \whs.   

 % ------------------------------------------------ fig 9 
\begin{figure}[t]
\centering
\includegraphics[width=7cm]{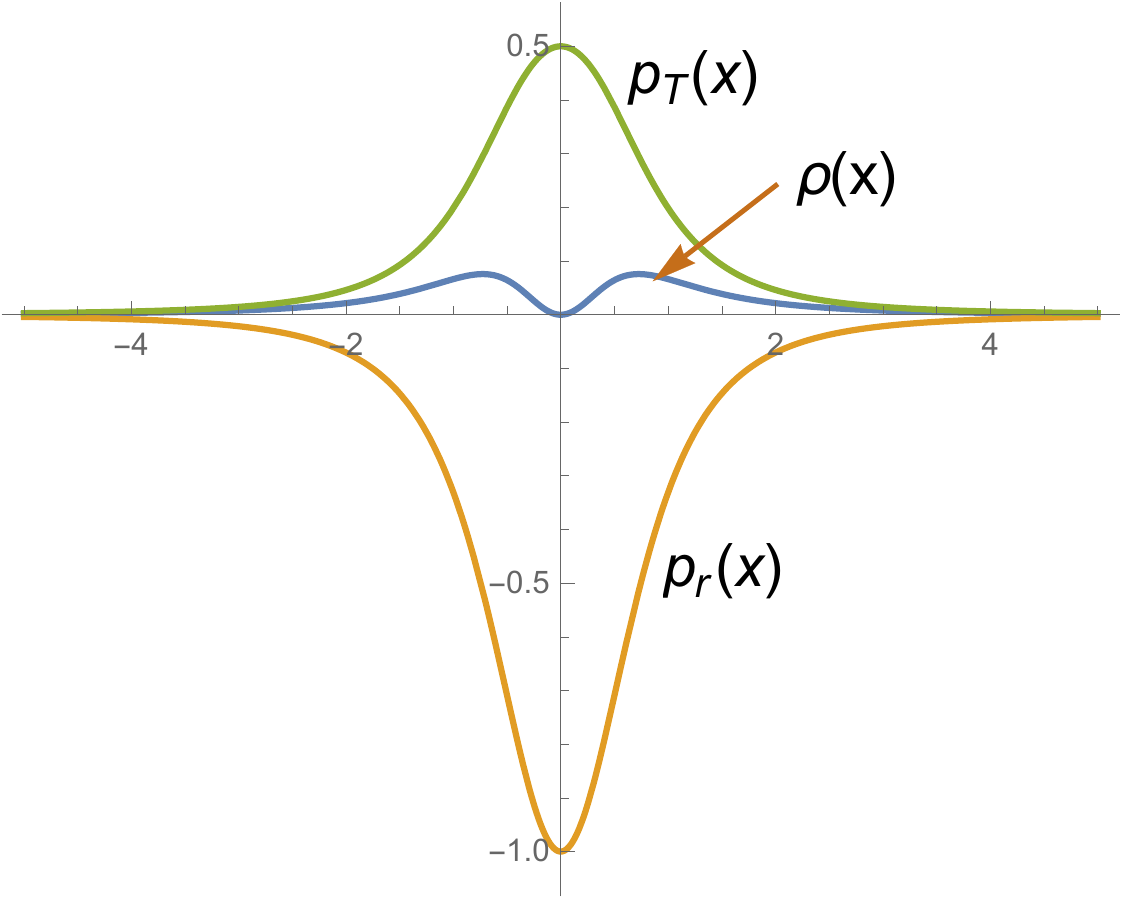} 
\caption{\small  
	The density $\rho(x)$ and the pressures $p_r(x)$ and $p_T(x)$ for a symmetric \wh\ with 
	$A(0) = 0.5$ and $A'(0) =0$.} \label{rho-sym}
\end{figure}
 % ------------------------------------------------ fig 10
\begin{figure}[ht]
\centering
\includegraphics[width=7cm]{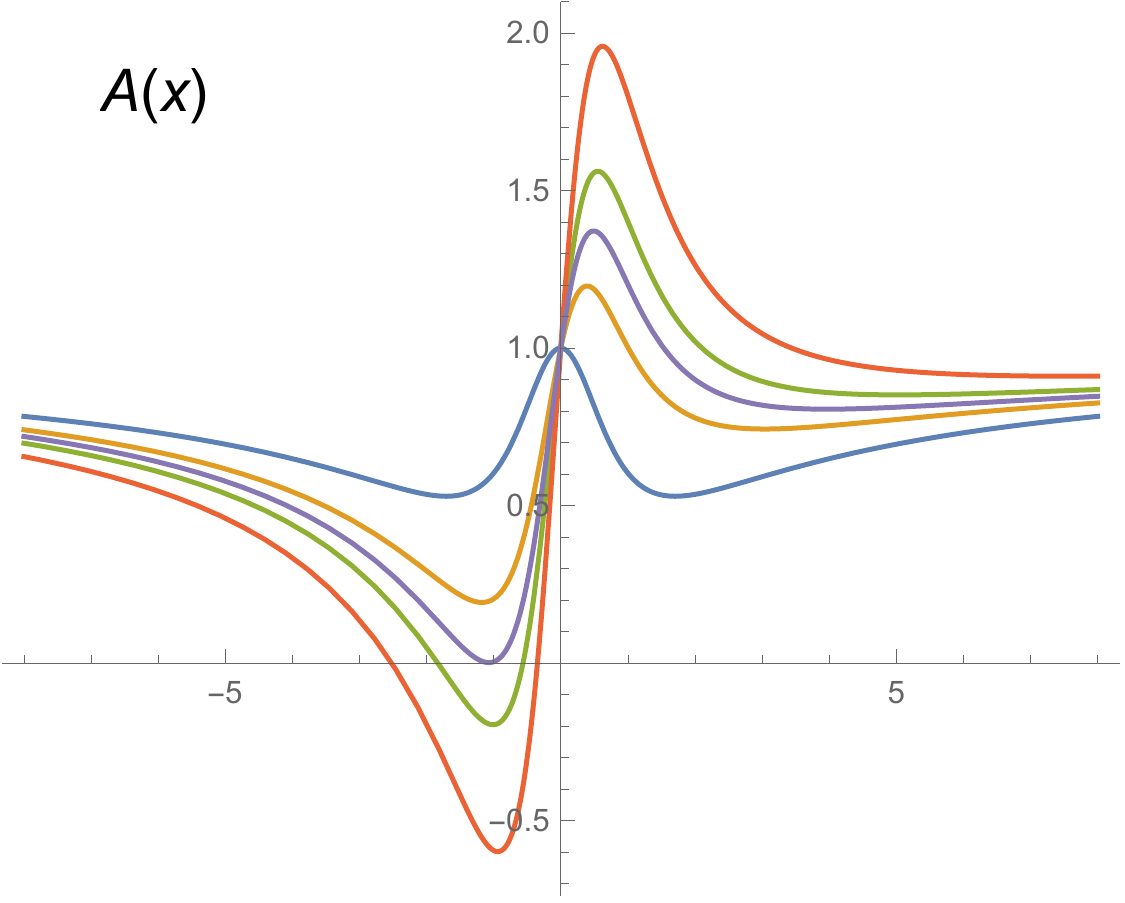} 
\caption{\small  
	Solutions to \eqn{eq-R0} with $A(0) =1$ and $A'(0) =0, 1, 1.5, 2, 3$ (upside-down for $x <0$ and 
          bottom-up for $x >0$).} 
\label{Aa-asym1}
\end{figure}
% -------------------------------------------------

  The effective matter density and pressures for an example of a symmetric \wh\ model with $A(0)=0.5$ and 
  $A'(0) =0$ are shown in Fig.\,\ref{rho-sym}. The NEC violation at all $x$ is evident since $\rho+p_r <0$. 

  The above pictures are weakly or strongly deformed if we specify $A'(0) \ne 0$. 
  Let us consider some examples. 

  The behavior of $A(x)$ in the case $A(0) =1$ is depicted in Fig.\,\ref{Aa-asym1}. One sees that 
  small  nonzero values of $A'(0)$ make the \wh\ asymmetric without changing its global structure, 
  but at $A'(0) \approx 1.5$ emerges a double horizon which turns into a pair of simple horizons at 
  larger $A'(0)$. A similar behavior of the
  solutions is observed for all $A(0) \lesssim 4.2$, at which the corresponding symmetric solutions
  have no zeros and describe \whs. 
 % ------------------------------------------------ fig 11
\begin{figure}[ht]
\centering
\includegraphics[width=6.5cm]{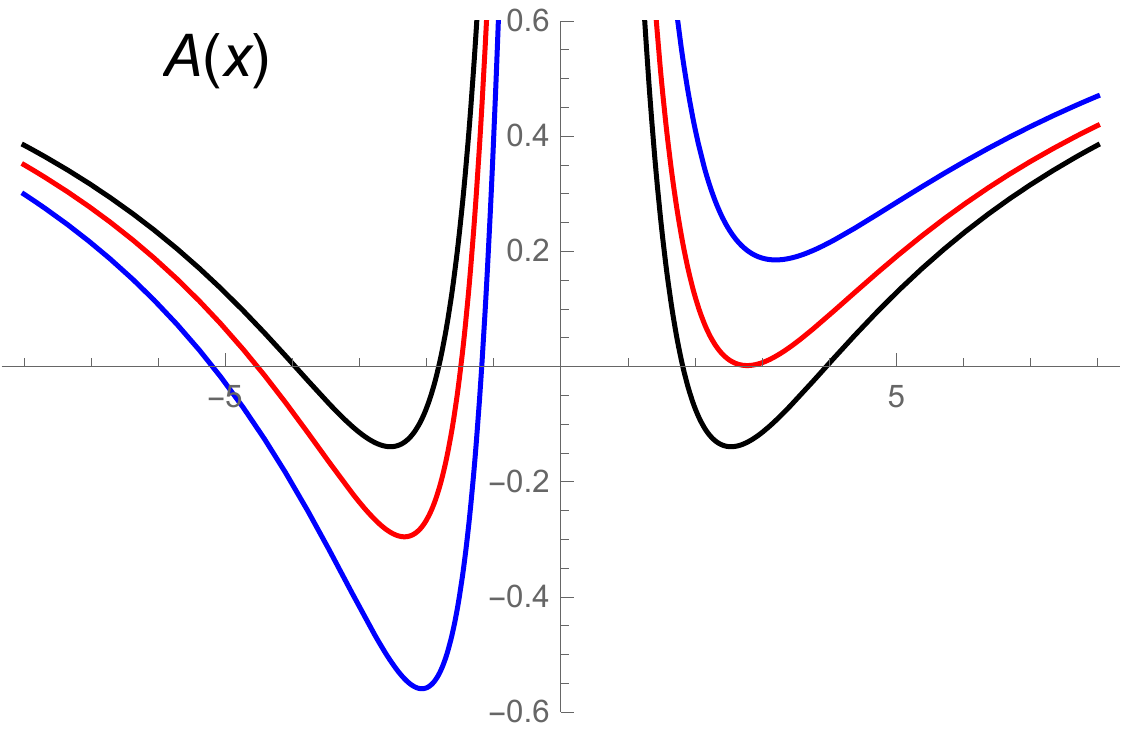} 
\caption{\small  
	Solutions to \eqn{eq-R0} with $A(0) =5$ and $A'(0) =0, 0.8, 2$ (upside-down for $x <0$ and 
          bottom-up for $x >0$). The peaks near $x=0$ are similar to those in Fig.\,\ref{Aa-sym}.} 
 \label{Aa-asym2}
\end{figure}
 % ------------------------------------------------ fig 12
\begin{figure}[ht]
\centering
\includegraphics[width=7cm]{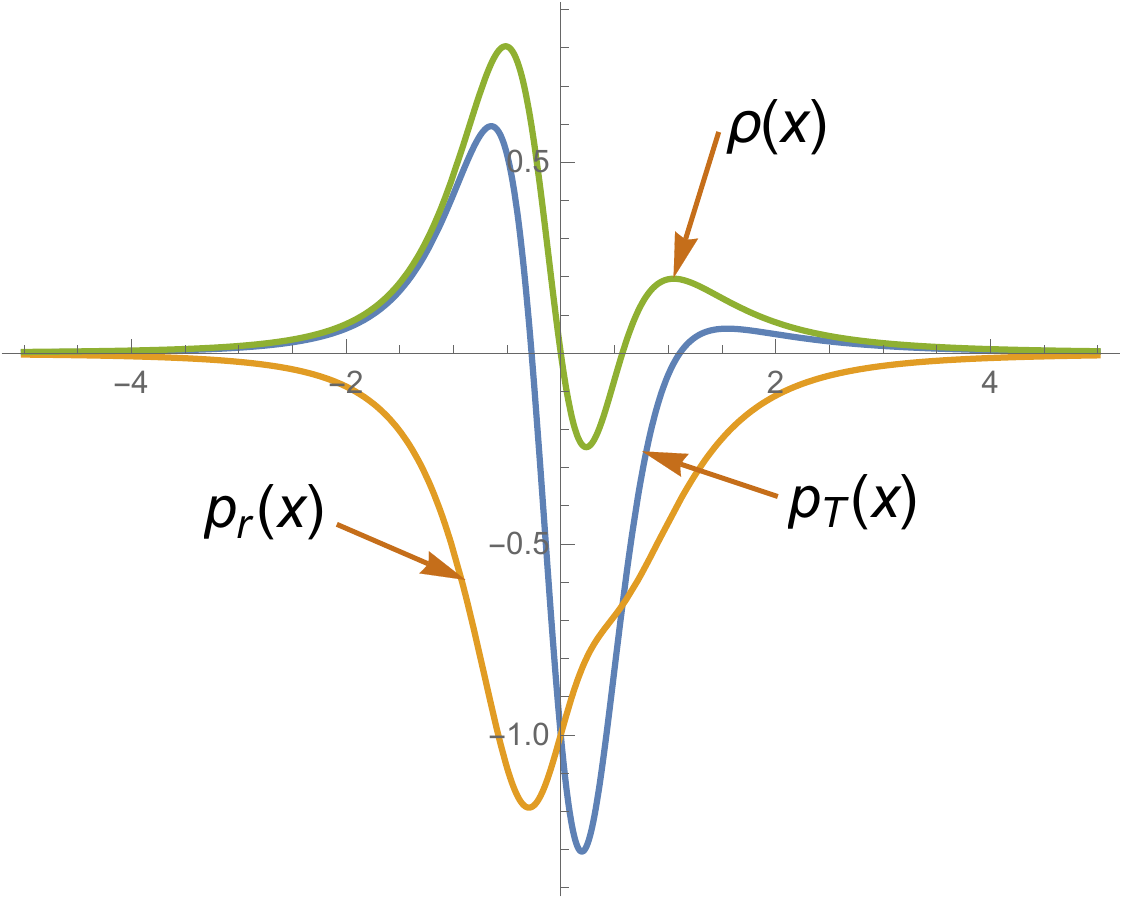} 
\caption{\small  
	The density $\rho(x)$ and the pressures $p_r(x)$ and $p_T(x)$ for an asymmetric \wh\ with 
	$A(0) = 1$ and $A'(0) =1$.} \label{rho-asym}
\end{figure}
 % ------------------------------------------------ 
  At larger values of $A(0)$, at which even functions $A(x)$
  have zeros shown in Fig.\,\ref{Aa-sym} (they describe symmetric regular \bhs), the
  corresponding asymmetric space-times are also regular \bhs, with the number of horizons from 
  two to four, as is evident from Fig.\,\ref{Aa-asym2}, showing $A(x)$ with $A(0) = 5$ and 
  different $A'(0)$. 
  
  Fig.\,\ref{rho-asym} presents an example of the behavior of the effective density and pressures in an 
  asymmetric \wh\ model. 

\section{Wormhole traversability and lensing}

  In this section we will briefly discuss some important properties of \asflat\ \whs\ taking as examples some
  of the $\Z_2$-symmetric solutions for which the redshift function $A(x)$ is plotted in Fig.\,8. The 
  corresponding numerical estimates will evidently be true by order of magnitude for other typical solutions. 

\subsection{Traversability}

\def\obet{\overline{\beta}{}}
\def\ogam{\overline{\gamma}{}}

  Not all \whs\ traversable by definition may really be used by a human being, or are ``traversable in
  practice'' \cite {viss-book}. A natural criterion for such traversability is that tidal accelerations due to 
  inhomogeneity of the gravitational field should not exceed the Earth's surface gravity, 
  $g_\oplus \approx 9.8\, {\rm m/s^2}$. For a body moving in the radial direction, the tidal accelerations 
  in a \ssph\ metric can be described by \eqs (13.4) and (13.6) from \cite{viss-book}, which can be 
  rewritten  as follows in the notations of the metric \rf{ds-q}:            
\bearr                      \label{acc1}
            \Delta a_{\|} = R^{tx}_{\ \ tx}\Delta\xi_{\|} =  \Half A'' \Delta\xi_{\|},
\yyy 				\label{acc2}
      \Delta a_{\bot} =\ogam^2 (R^{t\theta}_{\ \ t\theta} -\obet^2 R^{x\theta}_{\ \ x\theta}\Delta\xi_{\bot})
\nnn   \qquad
	    = \ogam^2 \biggl[- \Half \frac{A' r'}{r} 
			  + \obet^2\biggl(\frac{Ar''}{r} + \Half \frac{A'r'}{r}\biggr)\biggr]\Delta\xi_{\bot},
\ear
  where $\Delta a$ are the tidal accelerations in the radial (${}_{\|}$) and transveral (${}_{\bot}$) directions,
  and $\Delta\xi$ are small displacements in the same directions.\footnote
	{We have replaced the tetrad components of the Riemann tensor used in Visser's book \cite{viss-book}
	  with the mixed components $R^{\mu\nu}_{\ \ \mu\nu}$ (no summing). A direct calculation shows that 
 	  this replacement is equivalent due to diagonality of the metric \rf{ds-q} and diagonality of the
               Riemann tensor with respect to pairs of indices in the same metric, one should only take into account
	  the sign changes when raising the indices and the Riemann tensor definition in \cite{viss-book}.} 
  Furthermore, $\obet = v/c$ is the velocity (in units of the speed of light) relative to the static reference
  frame, and $\ogam = (1-\obet^2)^{-1/2}$ is the corresponding Lorentz factor. The expression \rf{acc2} is 
  especially simple for the throat $x=0$: since $r'(0) =0$, it follows 
\beq 				\label{acc3}
		\Delta a_{\bot} = \obet^2\ogam^2 \frac{A r''}{r} \Delta\xi_{\bot}.
\eeq
  In numerical estimates, the dimensionless quantities involved ($A, A'', r, r'', \obet, \ogam$) are of the
  order of unity provided that $a=1$. Indeed, by \rf{r_x}, at the throat $x=0$ we have $r=1$ and $r''=1$, 
  and by \rf{eq-R0} $A''(0) = 2 - 4A(0)$, the initial data for $A$ and $A'$ are taken to be of the 
  order of unity, and the values of all relevant quantities at $x\ne 0$ are of the same order as at $x=0$ 
  or smaller.

  However, our unit length is the arbitrary length $a$ equal to the throat radius, hence to obtain 
  estimates in meters, each $d/dx$ should be divided by $a$ expressed in meters. We should also take 
  into account that in the units where $c=1$ a second equals $3 \ten{8}$ m, therefore
  $g_\oplus = 9.8\, {\rm m/s^2}\approx 1.1\ten{-16}\, {\rm m}^{-1}$. 
  So, assuming $\Delta\xi_{\|, \bot} = 2$ m in \eqs \rf{acc1}--\rf{acc3}, 
  from the requirement $|\Delta a_{\|, \bot}| \lesssim g_\oplus$ we obtain
\beq                             \label{acc4}
                        a \gtrsim 10^8\ {\rm m} = 10^5\ {\rm km},  
\eeq 
  that is, the throat radius must be larger than roughly eight Earth's diameters. One can notice that 
  the estimated tidal accelerations at a \wh\ throat are of the same order of magnitude as 
  tidal accelerations at a \Scz\ horizon of  the same radius (see, e.g., \eqs (13.17) and (13.18) in 
  \cite{viss-book}).
  
  Another requirement is that a traveler should not experience too large center-of-mass accelerations.
  However, if the spacecraft moves along a geodesic, such an acceleration is zero (the usual
  free-fall weightlessness), and if not, everything depends of the engine activity. 

% -----------------------------------------------------
\subsection{Gravitational lensing}

  To calculate gravitational lensing as one of the most important potentially observable effects of 
  \whs, one can use the general formulas for \asflat\ \ssph\ space-times \cite{bozza-02}
  (see, e.g., \cite{tsu-16} for more recent references with calculations of light bending in 
   various \wh\ models). In our notations, the deflection angle $\alpha$, found by considering 
  null geodesics in the metric \rf{ds-q}, is given by 
\bearr
	\alpha = \alpha(x_0) = I(x_0) - \pi,	
\nnn								\label{lens1}
          I(x_0) = 2 \int_{x_0}^\infty \frac {dx}{r(x) \sqrt {r^2(x)/b^2 - A(x)} },
\ear 
  where $b = L/E$ is the so-called impact parameter charactering a particular null geodesic with 
  the conserved energy parameter $E = A(x) dt/d\sigma$ and the conserved angular momentum 
  $L = d\varphi/d\sigma$, $\sigma$ being an affine parameter along the geodesic. The coordinate
  value $x_0$ corresponds to the nearest approach of the photon path to the throat and is found 
  from the condition $dr/d\sigma =0$ which leads to
\beq 			\label{lens2}
	A(x_0) b^2 = r^2 (x_0).
\eeq
  It is clear that if $\cR := r(x_0) > 1$, then the photon is scattered against the \wh, while the equality
  $\cR = 1$ tells us that the photon reaches the throat and then passes through the \wh. We will restrict
  our calculations to photon paths with $\cR > 1$, mentioning that paths traversing a \wh\ 
  and images of another universe thus observed were considered in \cite {sha-09, sha-15}. Let us select 
  for consideration three particular $\Z_2$-symmetric models from those depicted in Fig.\,8, namely,
  one with $A(0) = 1/3$ (in which $A(x)$ has a minimum on the throat), another with $A(0)=1$
  (in which  $A(x)$ is slightly peaked at the throat such that $A(0) = A(\infty)$), and the third one 
  with $A(0) = 3$ with a larger peak of $A(0)$ at the throat. A numerical calculation leads to the 
  results shown in Fig.\,13, covering a range of $x$ from a close vicinity of the throat to those where 
  the deflection angles actually begin to follow the asymptotic law according to Einstein's formula
\beq                                        \label{lens-asy}
                    \alpha (\cR) \approx 4m/\cR,       
\eeq
  where $m$ is the \Scz\ mass that characterizes the almost Newtonian gravitational field at large $x$. 
  As before, the scale along the horizontal axis corresponds to $a = 1$, that is, the radius $r(x)$ is
  shown in units of the throat radius. Thus we compare light deflection near different
  \whs\ with the same throat radius. These \whs\ have different \Scz\ masses $m(A(0))$:   
\beq                                          \label{lens4}
                   m(0.3) \approx 0.89,\quad m(1) \approx 1.17, \quad  m(3) \approx 2.09
\eeq
  (recall that these are ``geometrized'' masses with the dimension of length, $m=GM$, $G$ being the 
  Newtonian gravitational constant and $M$ the conventional mass). For comparison, we also show the 
  deflection angles for an Ellis \wh\ \cite{br73, ellis} with the same throat radius, described by the metric  
  \rf{ds-q}  with $r^2 = 1+x^2$ and $A(x) \equiv 1$ (its lensing properties were analyzed in detail in, e.g., 
  \cite{abe-10, tsu-16a}), and for a \Scz\ \bh\ with the same horizon radius, $2m =1$ (see detailed 
  descriptions of \Scz\ \bh\ lensing in, e.g., \cite{darwin-59, virb-99, zupko-08}).     

 % ------------------------------------------------ fig 13
\begin{figure}[ht]
\centering  \hspace{-1cm}
\includegraphics[width=9cm]{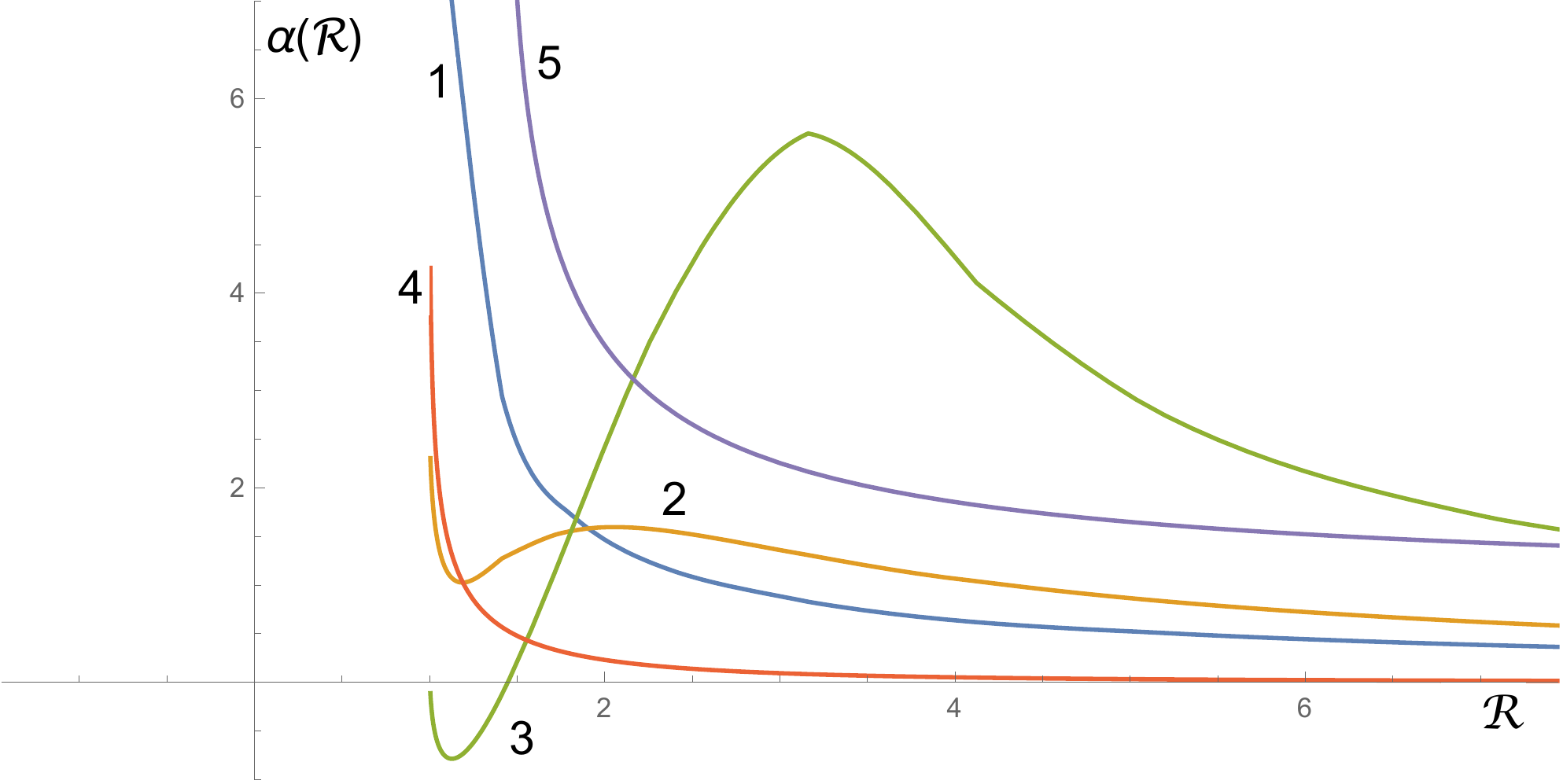} 
\caption{\small
        Light bending angles $\alpha$ as functions of the closest approach radius $\cR$ for 
	different \whs\ and a \Scz\ \bh. Curves 1--3 show $\alpha(\cR)$ for \whs\ with 
	the throat radius $a=1$ and $A(0) = 1/3, 1, 3$, respectively, curve 4 for an Ellis \wh\ 
	with the same throat radius, and curve 5 for a \Scz\ \bh\ with the same radius, $2m=1$,
        of the event horizon.
}  
\label{lens}
\end{figure}
 % ------------------------------------------------ 

  As is clear from the figure, the \wh\ lensing properties substantially depend on the profile of $A(x)$:
  curve 1, corresponding to $A(0) = 0.3$ such that $A(x)$ has a minimum at the throat, is rather similar to
  curve 5 describing \Scz\ \bh\ lensing, the main difference being that in the \Scz\ case  $\alpha \to \infty$ 
  as $\cR\to 3/2$ (a logarithmic divergence at the so-called photon sphere), while the same happens only 
  as $\cR\to 1$ for \whs.  Curve 2 shows a decrease in $\alpha$ in the region where $A(x)$ is decreasing. 
  Actually, the role of $A(x)$ is similar to that of the Newtonian gravitational potential in classical physics, so
  where $A' < 0$, the gravitational field is repulsive for both massive particles and photons. This effect is 
  still stronger if the peak of $A(x)$ is larger and can even lead to negative light bending at some $\cR$
  as can be seen for curve 3. Curve 4 pertains to the Ellis \wh, which is massless and therefore much 
  weaker affects the light beams, and its $\alpha(\cR)$ quicker decays at large $\cR$; 
  however, $\alpha(\cR)$ also diverges near $\cR=1$. 
  
 % ===================================
\section{Concluding remarks}
   
  Let us enumerate the main results of this study:

\begin{enumerate}
\item
   We have proved a no-go theorem showing that it is impossible to obtain static \asflat\ or AdS \whs\ 
   without horizons, supported by isotropic matter. It explains why in all previous attempts to build such 
   solutions it was necessary to introduce boundaries with thin shells. 
\item
  We have obtained a family of \whs\ with isotropic matter which connect two de Sitter worlds 
  with the same or different curvature. In the symmetric case, such ``bridges'' may connect distant
  regions of the same inflationary universe making them causally connected. It is 
  of interest that, unlike other models where the \wh\ throat expands together with the universes
  it connects, in our solutions the throat radius is constant. 
\item 
  It is important that even though we introduced, as sources of gravity, isotropic fluids in the static
  space-time region, these fluids inevitably become anisotropic in a T-region with a time-dependent
  Kantowski-Sachs type metric.
\item
   We have obtained a number of new numerical \asflat\ solutions to the equation $R = 0$,
   describing $\Z_2$-symmetric or asymmetric \wh\ and regular \bh\ configurations, among which 
   asymmetric ones are obtained quite naturally by specifying asymmetric initial data at $x=0$.
   Some \asflat\ metrics with $R=0$ contain up to four Killing horizons.
\item
   We have shown that the traversability condition for \whs\ considered here in terms of sufficiently low 
   tidal forces are actually the same as in other models and require a throat radius of about $10^5$ km 
   or more. A brief consideration of the lensing properties of our twice \asflat\ \whs\ have revealed their 
   distinguishing features, but a more complete analysis of this important phenomenon is postponed 
   for future studies.  
\item
  While solving the equations with respect to $A(x) =  g_{tt}$, expressing the isotropy condition 
  $p_r = p_T$ in Section 3, or zero scalar curvature in Section 4, we have revealed intersection 
  points in families of integral curves, corresponding to different initial values of $A$ but the same 
  initial slope $A'$. It is a manifestation of a general interesting property of linear ordinary differential
  equations, discussed in the Appendix. For the integral curves of $A(x)$, the existence of such
  intersections leads to the following general rule: given a fixed initial slope $A'(0)$, a curve that 
  begins higher at $x=0$ (that is, $A(0)$ is larger), ends lower at large $|x|$ in both positive 
  and negative directions.
\end{enumerate}

  Our results have been obtained in terms of the quasiglobal coordinate $x$. Let us comment on 
  some other choices of the radial coordinate which seem to be more intuitively understandable. 
  One of them is the curvature, or \Scz\ coordinate, $r$, equal to the spherical radius $\e^\beta$ 
  in the metric \rf{ds}. As already explained in the introduction, this choice is not good if $r$ 
  has a minimum, in particular, in all  \wh\ space-times. We can easily transform our solutions to $r$ 
  as a new coordinate by the substitution $x = \pm \sqrt{r^2 -1}$, which will result in two 
  separate branches for positive and negative $x$ in each solution. These branches will be identical 
  when obtained from $\Z_2$-symmetric solutions, but in asymmetric (that is, more general) ones the 
  unity of two branches will become quite non-obvious. However, $r$ is a convenient parameter for 
  showing the \wh\ lensing properties and their comparison with lensing by a \Scz\ \bh, see Fig.\,13. 

  Another popular choice is the Gaussian, or proper radial distance coordinate $l$, such that
  $\e^\alpha\equiv 0$ in \rf{ds}. This coordinate is quite suitable for describing \wh\ space-times
  but is not good enough for \bhs\ since at an extremal (double) horizon, where, in terms of 
  our metric \rf{ds-q}, $A(x) \sim (x - x_{\rm hor})^2$, the proper radial distance diverges. 
  So, if we used the coordinate $l$ for finding the solutions, we would lose their natural sequence 
  at transitions from \wh\ to \bh\ cases, and we would simply lose the solution with $A(0) =0$, 
  see Fig.\,8.     

  In a flat asymptotic region all three coordinates coincide, and at an (A)dS infinity the \Scz\ ($r$)
  and quasiglobal ($x$) coordinates also coincide. However, in a strong field region, as we see,
  the coordinate $x$ is the most preferable. It is always admissible at \wh\ throats, while at 
  horizons in \ssph\ space-times it is always finite and behaves, up to a nonzero constant factor,
  like a Kruskal-like coordinate needed to cross the horizon \cite{BR-book}; it can therefore 
  be used to jointly describe inner and outer regions of \bhs\ (hence the name ``quasiglobal''). 
  Apart from the fact that, in our notations, $r \equiv x$ in the \Scz-(A)dS solutions and their 
  charged counterparts, the coordinate $x$ is widely used in solutions with scalar fields, see, 
  e.g., \cite{almaz2, BBS12, bu1, GGS, kb-kor} and many others.

  It should be noted that such physically meaningful quantities as $A(x) \equiv g_{tt}$ and the nonzero 
  SET components $T\mN$ in the metric \rf{ds} (the density and pressures) are insensitive to the 
  choice of a radial coordinate and behave as scalars at its transformations. Therefore in cases 
  where two coordinates are equally admissible, such as $x$ and $l$ in \asflat\ \whs, a transition 
  from one such coordinate to another will merely result in non-uniform but finite stretching or 
  squeezing of the corresponding plots along the horizontal axis.  

% =====================
\section*{Appendix}
\def\theequation{A.\arabic{equation}}
\sequ 0

 In this Appendix we prove and discuss a very simple and interesting property of linear second-order 
 ordinary differential equations (L2-ODE), which must have numerous applications and must 
 probably be well known to mathematicians, but we were unable to find proper references.   

 Consider a general L2-ODE for $y(x)$:
\beq          \label{*1}
        A(x) y'' + B(x) y' + C(x) y = F(x),
\eeq
  where the prime means $d/dx$, and all quantities involved are supposed to be real.  
  Let there be initial conditions at some $x_0$:
\beq                        \label{*2}
	y(x_0) = a, \qquad y'(x_0) = b.
\eeq

  The general solution to \eqn{*1} is a function of $x$ and the initial data $a, b$. On the other hand,
\bearr      		\label{*3}
	y = y(x, a, b) = C_1 y_1(x) + C_2 y_2(x) + y_3(x),   
\nnn  
         C_1, C_2 = \const,
\ear
  where $y_1(x)$ and $y_2(x)$ are two linearly independent solutions to the homogeneous equation
  \rf{*1}, i.e., for $F(x)\equiv 0$, and $y_3(x)$ is a special solution to the inhomogeneous 
  equation \rf{*1}. 

  Suppose we know the functions $y_i (x)$.  
  Then, comparing \rf{*2} and \rf{*3}, we can write for $x=x_0$
\bearr                      \label{*4}
          C_1 y_{10} + C_2 y_{20}  = a - y_{30},
\nnn
          C_1 y'_{10} + C_2 y'_{20}  = b - y'_{30},
\ear
  with the constants $y_{i0} = y_i(x_0)$ and $y'_{i0} = y'_i(x_0)$, $i = 1,2,3$.
  The algebraic equations \rf{*4} may be used to express the constants $C_{1,2}$ in terms of 
  the initial data $a, b$. By Kramer's formulas, we have ($i=1, 2$)
\beq                          \label{*5}
          C_i =  \frac{W_i}{W_0}, \qquad  W_0 = \begin{vmatrix}  y_{10} & y_{20}\\
 								             y'_{10} & y'_{20}\end{vmatrix},
\eeq  
  where $W_0 \ne 0$ is the Wronskian of $y_{1,2}$ at $x=x_0$, and $W_i$ are the determinants
  obtained from $W_0$ by replacing its $i$-th column with that of the r.h.s. of \rf{*4}. Thus we obtain
\bearr                       
	C_1 = W_0^{-1} [(a-y_{30})y'_{20} - (b-y'_{30})y_{20}],   
\nnn
        	C_2 = -W_0^{-1} [(a-y_{30})y'_{10} - (b-y'_{30})y_{10}], 
\earn 
  Substituting this into \rf{*3}, we present the solution with explicit dependence on the initial data 
  $a$ and $b$:
\bearr                                \label{*6}
           y(x, a, b) = \frac{y_1(x)}{W_0}[(a-y_{30})y'_{20} - (b-y'_{30})y_{20}] 
\nnn 
			- \frac{y_2(x)}{W_0}[(a-y_{30})y'_{10} - (b-y'_{30})y_{10}]  + y_3(x).
\ear

  Now we put the following question: Is there such a value of $x$, say, $x=x_*$, at which the
  function $y$ takes the same value for any choice of $a$ if $b$ is fixed? It will mean that all 
  integral curves $y(x)$, beginning at $x_0$ with the same slope $y'(x_0) = b$ but different starting 
  points $y(x_0) = a$, intersect at $x = x_*$.

  If such a value does exist, then at $x = x_*$ we should have $\d y/\d a =0$, where $y = y(x,a,b)$
  is given by \rf{*6}. Explicitly, the condition $\d y/\d a =0$ has the form
\beq                                   \label{*7}
	y_1(x_*) y'_{20} = y_2(x_*) y'_{10}.
\eeq
  It is an algebraic (in general, transcendental) equation with respect to $x_*$ if the functions 
  $y_{1,2}(x)$ are known. This equation may have any number of solutions, from zero to infinity.
  Anyway, it is clear that the existence of such intersection points is quite a general phenomenon.

  Some important observations are in order:
\begin{enumerate}
\item
	It is easily verified that \eqn{*7} is insensitive to a particular choice of two linearly 
  	independent solutions $y_{1,2}$ to the homogeneous equation \rf{*1}.
\item
	The value of $x_*$ is insensitive to the inhomogeneity $F(x)$ in \eqn{*1}. It only depends on
  	the left-hand side of \rf{*1} and on the choice of $x_0$.   
\item
	The value of $x_*$ does not depend on $b$. In other words, different sets of integral 
         curves beginning at $x=x_0$ at different ``heights'' $a$ but with the same slope $b$,
	intersect  at the same $x=x_*$ for all values of $b$ (though at a $b$-dependent height).
\end{enumerate}

{\bf  Example.} Consider the simplest L2-ODE
\beq           \label{*8}
            y'' + Ky = L, \cm K, L = \const.
\eeq
  Then, first of all, we can write the solution $y_3 = L/K$ of the inhomogeneous equation, which 
  has no effect on anything further on. 

  Next, if $K = k^2 > 0$, we can write 
\beq                     \label{*9}
       y_1(x) = \cos kx, \quad y_2(x) = \sin kx,
\eeq
  and if we choose $x_0 =0$, then \eqn{*7} gives $\cos kx_*=0\ \then\ x_* = \pi/2 + \pi n$,
  where $n \in \Z$, an infinite number of solutions, or an infinite number of intersection points of the
  integral curves along the $x$ axis. The same result is obtained if we take other $y_{1,2}$,
  for example, $y_{1,2} = \cos kx \pm \sin kx$.

  If the initial data are specified at another $x_0$, the intersection points are located at other $x$.
  For example, if $x_0 = \pi/(4k)$, then \eqn{*7} gives $kx_* = -\pi/4 + \pi n$, $n\in \Z$, again
  an infinite number of intersection points, but they are located at other $x_*$ than for $x_0=0$.

  Lastly, if $K = -k^2 <0$, then, instead of \rf{*9},  
\bearr
       y_1 = e^{kx}, \qquad y_2 = e^{-kx}, \quad {\rm or\ equivalently} 
\nnn
       y_1 =\cosh kx, \qquad   y_2 = \sinh kx.
\ear    
  If we choose $x_0 =0$, \eq \rf{*7} leads to $\cosh kx_* =0$, so there is no solution. Thus the 
  integral curves of \eqn{*8} with $K<0$, beginning at $x_0=0$ with the same slope, do not 
  intersect. The same result is obtained for any $x_0 \ne 0$.  

  This example illustrates the observation that the number of intersection points of integral curves
  may vary from zero to infinity.  

% =======================================

\subsection*{Acknowledgments}

  We thank Sergei Bolokhov for helpful discussions. 
  The work of KB was partly performed within the framework of the Center FRPP 
  supported by MEPhI Academic Excellence Project (contract No. 02.a03.21.0005, 27.08.2013).
  This publication was also supported by the Ministry of Education and Science of the Russian Federation
  (Agreement number 02.A03.21.0008) and by RFBR grant 16-02-00602.

\end{document}